\documentclass[10pt, conference]{IEEEtran}

\usepackage{cite}
\usepackage{amsmath,amssymb,amsfonts}
\usepackage{algorithmic}
\usepackage{graphicx}
\usepackage{amsfonts}
\usepackage{textcomp}
\usepackage{xcolor}
\usepackage{url}
\usepackage{soul}
\usepackage{algorithm}
\usepackage{caption}
\usepackage{fontawesome}
\usepackage{subcaption}
\usepackage{phonetic}
\usepackage{interval}
\usepackage{multicol}
\usepackage{multirow}
\usepackage{xspace}
\usepackage{pifont}
\usepackage{booktabs}
\usepackage{float}
    
\newcommand{\blockumulus}{\mbox{Blockumulus}\xspace}

\definecolor{foocolor}{HTML}{03A60D}

\definecolor{myblue}{HTML}{0026ff}

\definecolor{nickgreen}{HTML}{03A60D}
\definecolor{dgreen}{HTML}{03A60D}

\newcommand{\cmark}{\ding{51}}%
\newcommand{\xmark}{\ding{55}}%
    
\begin{document}

\title{Blockumulus: A Scalable Framework for Smart Contracts on the Cloud}

\author{\IEEEauthorblockN{
Nikolay Ivanov\textsuperscript{1}, 
Qiben Yan\textsuperscript{1},
Qingyang Wang\textsuperscript{2}
}
% %\\
 \IEEEauthorblockA{
 $^1$Computer Science \& Engineering, Michigan State University, East Lansing, MI, USA.}
  \IEEEauthorblockA{
 $^2$Computer Science \& Engineering, Louisiana State University, Baton Rouge, LA, USA.}
}

\maketitle

\begin{abstract}
Public blockchains have spurred the growing popularity of decentralized transactions and smart contracts, especially on the financial market. However, public blockchains exhibit their limitations on the transaction throughput, storage availability, and compute capacity. To avoid transaction gridlock, public blockchains impose large fees and per-block resource limits, making it difficult to accommodate the ever-growing high transaction demand. Previous research endeavors to improve the scalability and performance of blockchain through various technologies, such as side-chaining, sharding, secured off-chain computation, communication network optimizations, and efficient consensus protocols. However, these approaches have not attained a widespread adoption due to their inability in delivering a cloud-like performance, in terms of the scalability in transaction throughput, storage, and compute capacity. 

In this work, we determine that the major obstacle to public blockchain scalability is their underlying unstructured P2P networks. We further show that a centralized network can support the deployment of decentralized smart contracts. We propose a novel approach for achieving scalable decentralization: instead of trying to make blockchain scalable, we deliver decentralization to already scalable cloud by using an Ethereum smart contract. We introduce Blockumulus, a framework that can deploy decentralized cloud smart contract environments using a novel technique called overlay consensus. Through experiments, we demonstrate that Blockumulus is scalable in all three  dimensions: computation, data storage, and transaction throughput. Besides eliminating the current code execution and storage restrictions, Blockumulus delivers a transaction latency between 2 and 5 seconds under normal load. Moreover, the stress test of our prototype reveals the ability to execute 20,000 simultaneous transactions under 26 seconds, which is on par with the average throughput of worldwide credit card transactions.
\end{abstract}


\section{Introduction}\label{section:introduction}

Bitcoin is the first decentralized digital currency powered by blockchain with proof-of-work (PoW) consensus, which effectively prevents data tampering by anyone with less than half of the total computational power of the network~\cite{nakamoto2019bitcoin}. Recently, Dembo et al. delivered a formal proof of the correctness of the above statement with respect to the original PoW Nakamoto consensus~\cite{dembo2020everything}. Although Bitcoin's original purpose was to serve as a cryptocurrency transaction ledger, the unique properties of blockchain soon attracted researchers and engineers to re-purpose the technology for a plethora of decentralized applications, commencing the era of \emph{smart contracts}.

Smart contracts are decentralized immutable programs that allow to establish custom mediator-free protocols between parties that do not trust one another. For example, a smart contract can be used to help conduct an election in a decentralized manner~\cite{hardwick2018voting,wang2018large}. Another popular use case is \emph{fungible tokens}, which can represent corporate shares, gift card balances, and even custom currencies. Recently, researchers and businesses proposed a wide variety of smart contract applications~\cite{ritzdorf2018tls,ivanov2020smart,ramachandran2018smartprovenance}, some of which have already been adopted by nations' governments and large industries~\cite{williams2016estonia}. However, the unique features of blockchain and smart contracts come at a high price of mediocre performance and bounded scalability.

One way to address the performance and scalability issues of blockchain is to use a private permissioned blockchain framework, such as Hyperledger Fabric~\cite{androulaki2018hyperledger}, which only uses  pre-installed smart contracts (called chaincode) and splits the voting power between a small number of fixed participants. Although such blockchains deliver performance improvement over public blockchains, the requirement to establish a trustworthy consortium of organizations running these blockchains prevents its wide adoption in many applications, such as cryptocurrencies and decentralized voting. Thus, public blockchains cannot be replaced by permissioned ones.

%To address the inherent performance and scalability issues of public blockchains, 
Recently, a number of solutions have been developed to address the inherent performance and scalability issues of public blockchain, including partial off-chain computation~\cite{cheng2019ekiden}, side-chaining~\cite{poon2017plasma}, cross-chaining~\cite{wood2016polkadot}, sharding~\cite{zamani2018rapidchain,gilad2017algorand}, payment channels~\cite{poon2016bitcoin,dziembowski2017perun}, efficient consensus protocols~\cite{biswas2019pobt}, new blockchain architecture~\cite{sompolinsky2016spectre}, and network optimizations~\cite{chawla2019velocity}. However, all these solutions suffer from at least one of the following limitations. First, they could not deliver scalability in transaction throughput, data storage, and computation capacity at the same time. Second, the performance improvement is often incremental, but could be insufficient for many applications, such as retail payments. Third, they either do not support smart contracts, or their smart contracts are not Turing-complete~\cite{minsky1967computation}, making it impossible to realize certain programming patterns. 
%rendering some crucial programming patterns impossible, e.g., loop breaks determined at runtime\footnote{cf. Turing-complete Ethereum smart contracts~\cite{wood2014ethereum} and Turing-incomplete Bitcoin scripts~\cite{script}.}. 
%frameworks have significant limitations 
%\commentyan{what do you mean by ``significant limitations"?}\done. 
A recent blockchain scalability survey by Zhou et al.~\cite{zhou2020solutions} concludes that a desired solution still has not been found. In this paper, we propose a conceptually new approach to the blockchain scalability problem: \emph{we use an existing blockchain as-is to enable smart contracts on an already scalable system: the cloud.}

\noindent\textbf{Observation 1: Centralized Service for Scalable Decentralized Contracts.}
The operation of decentralized systems is often supported by underlying centralized and/or permissioned services. For example, the decentralization of Domain Name System (DNS) is based on the assumption that the Internet Corporation for Assigned Names and Numbers (ICANN), which oversees the system, is functional and trustworthy~\cite{rfc2870}. Such pattern is also observed in public blockchains. Kwon et al.~\cite{kwon2019impossibility} formally demonstrate that classic public blockchains exhibit partial centralization incurred by concentration of compute power around a few mining pools. Moreover, the decentralized nodes of blockchains use the Internet as a communication medium, which is subsequently enabled by a network of centralized routers and Internet service providers (ISPs), whose owners must comply with the regulations of local and federal jurisdictions. In this work, we extrapolate the above principle (i.e., the reconciliation of centralization and decentralization) to show that it is feasible and beneficial to build an environment that uses a centralized cloud as an underlying communication, storage and compute service for decentralized smart contracts. Particularly, \emph{this work demonstrates that cloud resources, such as storage and computation, can be treated as a utility (offered by a third party), which can support the operation of a decentralized network.}

\noindent\textbf{Observation 2: High Cost of Permissionless Network.} Public blockchains are supported by unstructured permissionless P2P networks, where nodes can freely join and leave. To support such a flexibility, the blockchains use a gossip protocol for peer communication. In this protocol, the peers are unaware of the current configuration of the network, so they achieve the network-wide propagation of broadcast messages by forwarding them through a subset of known peers. This incurs a significant message propagation latency and strict limits on the amount of data that can be transferred~\cite{decker2013information,zamani2018rapidchain}. Moreover, to prevent Sybil attacks, in which an adversary creates a large number of fake identities for gaining greater voting power, the PoW consensus algorithm has been used by Bitcoin, Ethereum, and many other popular blockchains. The PoW consensus involves a heavy computation, resulting in enormous electricity consumption.
%, for Bitcoin only, greater than used by entire Austria~\cite{zhou2020solutions}. 
As such, public blockchains pay a very high price for the flexibility of the underlying P2P network. 
% \commentyan{The following sentence needs to be rewritten as pointed out by Reviewer 3.}\done
%\emph{In this work, we separate the concept of decentralized contracts from the concept of decentralized network, and further show that decentralized smart contracts can be sustained by more efficient centralized network.}
\emph{In this work, we show that a smart contract can be used to facilitate a decentralized consensus in an overlay smart contract environment built upon a centralized network of cloud providers, which drastically reduces communication and computational overhead.}

Putting together the above observations, we develop the concept of \emph{overlay consensus}, which aims to deliver decentralization to smart contracts in a centralized cloud instead of random P2P network nodes. As a result, a consortium of clouds can host a permissionless smart contract environment and sell the access to it, but it cannot control the execution of these contracts or interfere with the data stored by these contracts. To achieve this, we use a smart contract deployed on a public blockchain to accrue periodic proofs of decentralization reported by the cloud consortium. The smart contract is designed in a way that any attempt of a foul play would inevitably generate a publicly-verifiable  proof for the action of breaking the consensus protocol.

In summary, we make the following contributions:

\begin{itemize}
    \item We introduce \blockumulus\footnote{The name \blockumulus is the portmanteau of the words ``blockchain'' and ``cumulus'' --- a type of cloud with the traditional puffy texture.}, a distributed framework for cloud smart contracts (bContracts\footnote{bContract stands for ``\blockumulus contract''.}) based on the novel concept of overlay consensus.
    
    \item We implement the full \blockumulus stack along with a sample bContract, called \emph{FastMoney}, for payment processing.
    
    \item We evaluate our \blockumulus implementation and the \emph{FastMoney} bContract to show that the framework delivers low transaction latency, high transaction throughput, and affordable operation cost. 
    
    %\item In the spirit of open research, we make the source code of \blockumulus and evaluation data publicly available\footnote{\url{https://github.com/nick-ivanov/blockumulus}}.
\end{itemize}

\section{Related Work}\label{section:relatedworks}

\begin{table}[]
    \centering
    \caption{\textbf{Comparison of \blockumulus with state-of-the art solutions.
    %addressing blockchain scalability and performance issues.
    }
    %We exclude some predecessor works that have been outperformed by works listed in this table.
    }
    \label{tab:comparison}
    \begin{tabular}{|c|c|c|c|c|}
    \hline
         \multirow{2}{*}{\textbf{Solution}} & \scriptsize{\textbf{General-purpose smart}} & \multicolumn{3}{c|}{\textbf{Scalability improvement}} \\
         \cline{3-5}
         & \scriptsize{\textbf{contract support}} & \scriptsize{\textbf{TPS$^{\mathrm{a}}$}} & \scriptsize{\textbf{Storage}} & \scriptsize{\textbf{Compute}} \\
    \hline\hline
        Algorand~\cite{gilad2017algorand} & \xmark & \cmark & \xmark & \xmark \\
    \hline
        RapidChain~\cite{zamani2018rapidchain} & \xmark & \cmark & \xmark & \xmark \\
    \hline         
        %Bitcoin & & & & \\
        Lightning~\cite{poon2016bitcoin} & \xmark & \cmark & \xmark & \xmark \\
    \hline
        Ekiden~\cite{cheng2019ekiden} & \cmark & \cmark & \xmark & \cmark \\
    
    % \hline
    %     Hyperledger Fabric~\cite{androulaki2018hyperledger} & \xmark & \cmark & \cmark & \cmark \\
    % \hline
    %     Raiden~\cite{raiden} & \xmark & \cmark & \xmark & \xmark \\
    % \hline
    %     Sprites~\cite{miller2019sprites} & \xmark & \cmark & \xmark & \xmark \\
    % \hline 
    %     Duplex Micropayment & & & & \\
    %     Channels~\cite{decker2015fast} & & & & \\
    
    % \hline
    %     TrueBit~\cite{teutsch2019scalable} & \xmark & \xmark & \xmark & \cmark \\

    \hline
        Arbitrum~\cite{kalodner2018arbitrum} & \xmark & \xmark & \xmark & \cmark \\
        
    \hline
        Jidar~\cite{dai2019jidar} & \xmark & \xmark & \cmark & \xmark \\

    % \hline
    %     ELASTICO~\cite{luu2016secure} & \xmark & \cmark & \xmark & \xmark \\
        
    \hline
        Monoxide~\cite{wang2019monoxide} & \xmark & \cmark & \cmark & \xmark \\

    % \hline
    %     Kadcast~\cite{rohrer2019kadcast} & \cmark & \cmark & \xmark & \xmark \\

    \hline
        Plasma~\cite{poon2017plasma} & \cmark & \xmark & \faQuestion & \cmark \\

    % \hline
    %     Velocity~\cite{chawla2019velocity} & \faQuestion & \cmark & \xmark & \xmark \\

    % \hline
    %     SPECTRE~\cite{sompolinsky2016spectre} & \xmark & \cmark & \xmark & \xmark \\
    
    \hline
        OmniLedger~\cite{kokoris2018omniledger} & \xmark & \cmark & \xmark & \xmark \\
    
    \hline
    
        \textbf{\blockumulus} & \cmark & \cmark & \cmark & \cmark \\
    \hline

    \multicolumn{5}{l}{$^{\mathrm{a}}$Transaction throughput (transactions per second).}
    \end{tabular}
    \vspace{-10pt}
\end{table}

Table~\ref{tab:comparison} compares the state-of-the art solutions aiming to address the blockchain scalability and performance limitations.
%The major limitation of all these systems is their failure to deliver a scalable environment for general-purpose smart contracts, i.e., the smart contract suitable for a variety of applications beyond cryptocurrency transactions.
Although these studies improve the blockchain scalability, they could not simultaneously accommodate the growing demand for transaction throughput, data storage, and heavy computation, in applications such as cryptography, AI, and big data analytics. In contrast, \blockumulus brings general-purpose smart contracts (i.e., the smart contract suitable for a variety of applications beyond cryptocurrency transactions) on the cloud, which improves blockchain scalability in terms of transaction throughput, data storage, and computation simultaneously.

\noindent\textbf{Off-chain Execution.}
Off-chain execution is an arrangement that allows to perform computation of some portions of smart contracts outside of the blockchain to improve performance and reduce costs. Ekiden~\cite{cheng2019ekiden} addresses the lack of confidentiality and poor performance of blockchain by securing an off-chain computation via trusted execution environment (TEE) technology. Despite significant performance improvement, the operation of Ekiden relies on the availability of crowdsourced consensus and compute nodes. The security of the system is founded on the assumption that the participants have Sybil-resistant identities (i.e., they cannot create multiple fake accounts). 
%, which may be unrealistic. 
The requirement for a participation deposit to prevent Sybil attacks may not only be ineffective against wealthy attackers, but may also reduce the incentive for community participation. Another off-chain execution solution, ZEXE, is proposed for abundant private off-chain computation~\cite{bowe2020zexe}. Unlike Ekiden, ZEXE does not require hardware TEE enclaves, and therefore can be used in a wider scope of platforms. However, this system focuses on improving the computation scalability and reducing communication overhead, whereas the scalability issue in storage and transaction throughput remains unaddressed.

\noindent\textbf{Side-Chaining and Cross-Chaining.}
Side-chaining is an arrangement in which some smart contract execution is outsourced to a different blockchain, while cross-chaining is a way for independent blockchains to share resources and use common assets. Plasma~\cite{poon2017plasma} attempts to reduce fees and improve performance of Ethereum blockchain by linking a smart contract to a tree of child blockchains. Although Plasma distributes the computation load of the master smart contract among multiple chains, the transaction throughput remains a likely bottleneck, and there is no solid evidence of significant improvement of storage capacity. A popular cross-chaining solution called Polkadot~\cite{wood2016polkadot} improves transaction throughput by creating a network of interoperable blockchains. However, the solution does not directly address the storage and compute capacity for smart contracts.

\noindent\textbf{Sharding and Alternative Consensus.}
The concept of sharding involves selecting a subset of nodes to serve as temporary representatives in a decentralized consensus, which curbs the performance degradation associated with gossip broadcasts in large blockchain networks. Algorand~\cite{gilad2017algorand} proposes a blockchain with improved performance using a sharding scheme based on a verifiable random function. Algorand delivers a significant increase in transaction throughput compared to classic public blockchains, but its operation relies on a set of assumptions that can be refuted by massive denial-of-service or Sybil attacks. Specifically, Algorand assumes that at least 95\% of all honest users must be able to send messages to other honest users, and the overall share of honest participants must be greater than 2/3. Another solution with sharding-based consensus is Rapidchain~\cite{zamani2018rapidchain}, which delivers high transaction throughput. However, Rapidchain is not scalable in terms of data storage and compute capacity.

Some alternative consensus models attempt to replace a compute-heavy PoW algorithm with lightweight alternatives, such as proof-of-stake (PoS), in which the voting power is determined by the amount of funds in possession of a node. Ouroboros~\cite{kiayias2017ouroboros} is a provably secure blockchain with PoS consensus. Unfortunately, existing alternative consensuses fail to address the full spectrum of scalability problems, and they introduce a significant fairness challenges, such as ``monetary hegemony''. Yu et al.~\cite{yu2020ohie} propose a lightweight consensus protocol, OHIE, to improve blockchain scalability by leveraging a parallel execution of the Nakamoto consensus. Despite the improvement in transaction throughput and available bandwidth, the scalability of storage and computation is not considered in OHIE.

\noindent\textbf{Network Optimizations and Payment Channels.}
Off-chain payment channels have been proposed to improve performance and reduce fees associated with financial transactions. The Lightning Network protocol~\cite{poon2016bitcoin} allows to create off-chain micropayment channels. Perun~\cite{dziembowski2017perun} is another proposal of a payment channel that improves routing of transactions. Although off-chain payment channels have been adopted by real-world applications, they cannot serve as alternatives of public blockchains because of their specific focus (only for payment) and the necessity to orchestrate a network of crowdsourced participants.

\noindent\textbf{Alternative Architectures.}
Researchers have been re-thinking the architecture of blockchain in order to improve performance and scalability. SPECTRE~\cite{sompolinsky2016spectre} proposes a reorganization of a traditional Nakamoto blockchain into a directed acyclic graph (DAG). Although it improves the speed of transactions, it could not be used for general-purpose smart contracts that may require abundant data storage and heavy computation.

\section{System Design}\label{section:systemdesign}
In this section, we introduce the \blockumulus framework and its operation protocol.

\subsection{\blockumulus Overview}

\begin{figure}
    \vspace{5pt}
    \centering
    \includegraphics[width=0.8\linewidth]{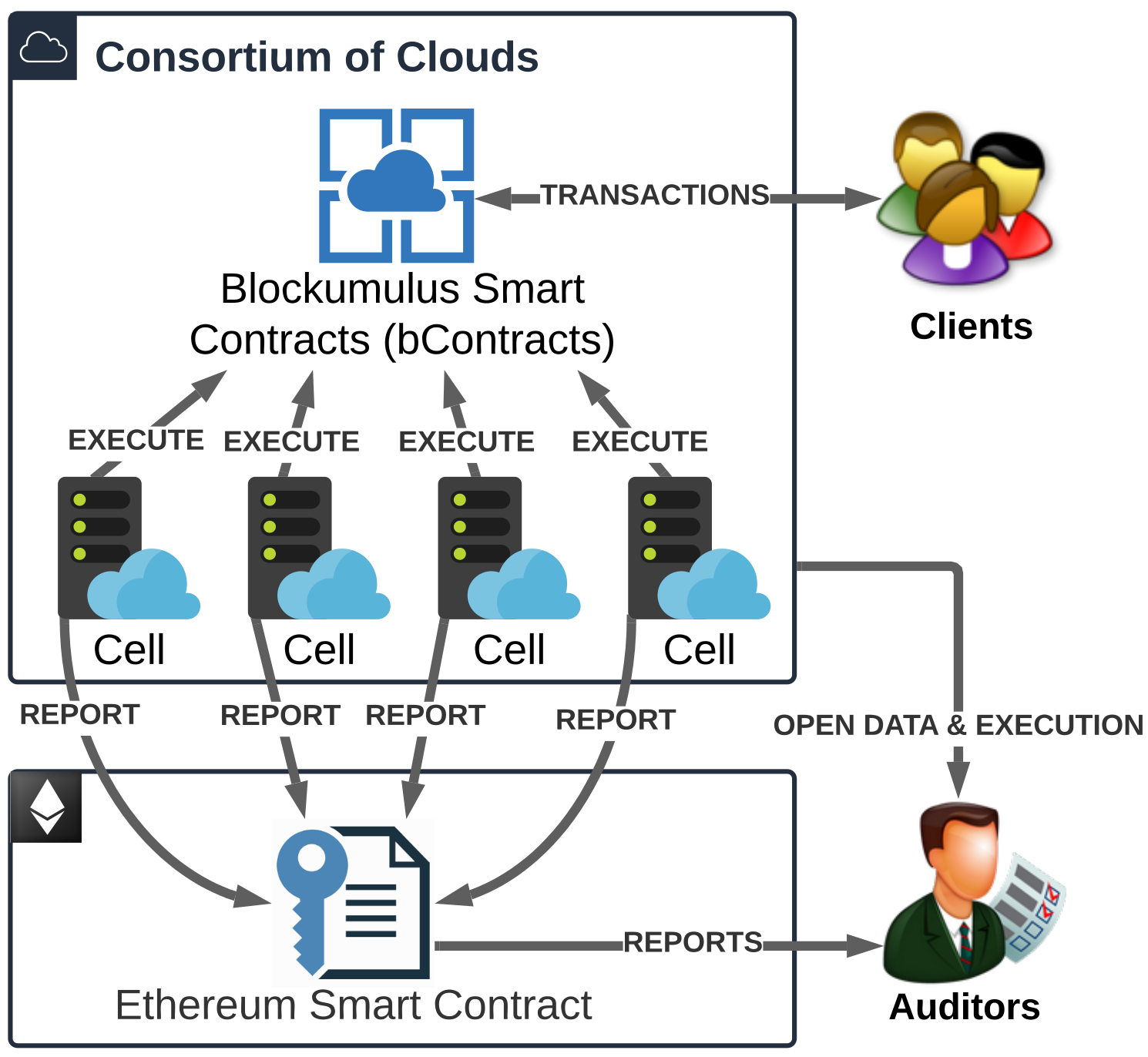}
    \caption{\textbf{\blockumulus overview.}
    }
    \label{fig:overview}
    \vspace{-10pt}
\end{figure}

\blockumulus is a framework that builds a decentralized environment for executing smart contracts upon a \emph{cloud consortium} --- a fixed set of $M$ cloud nodes called \emph{cells}, synchronized by the \emph{overlay consensus}. The overlay consensus is empowered by a smart contract deployed on a third-party public blockchain, with independent auditors running software for an automated verification of \blockumulus workflow (see Fig.~\ref{fig:overview}). Next, we introduce the major concepts of \blockumulus.

\subsubsection{\blockumulus Code Execution Model}
\begin{figure}
    \centering
    \includegraphics[width=0.55\linewidth]{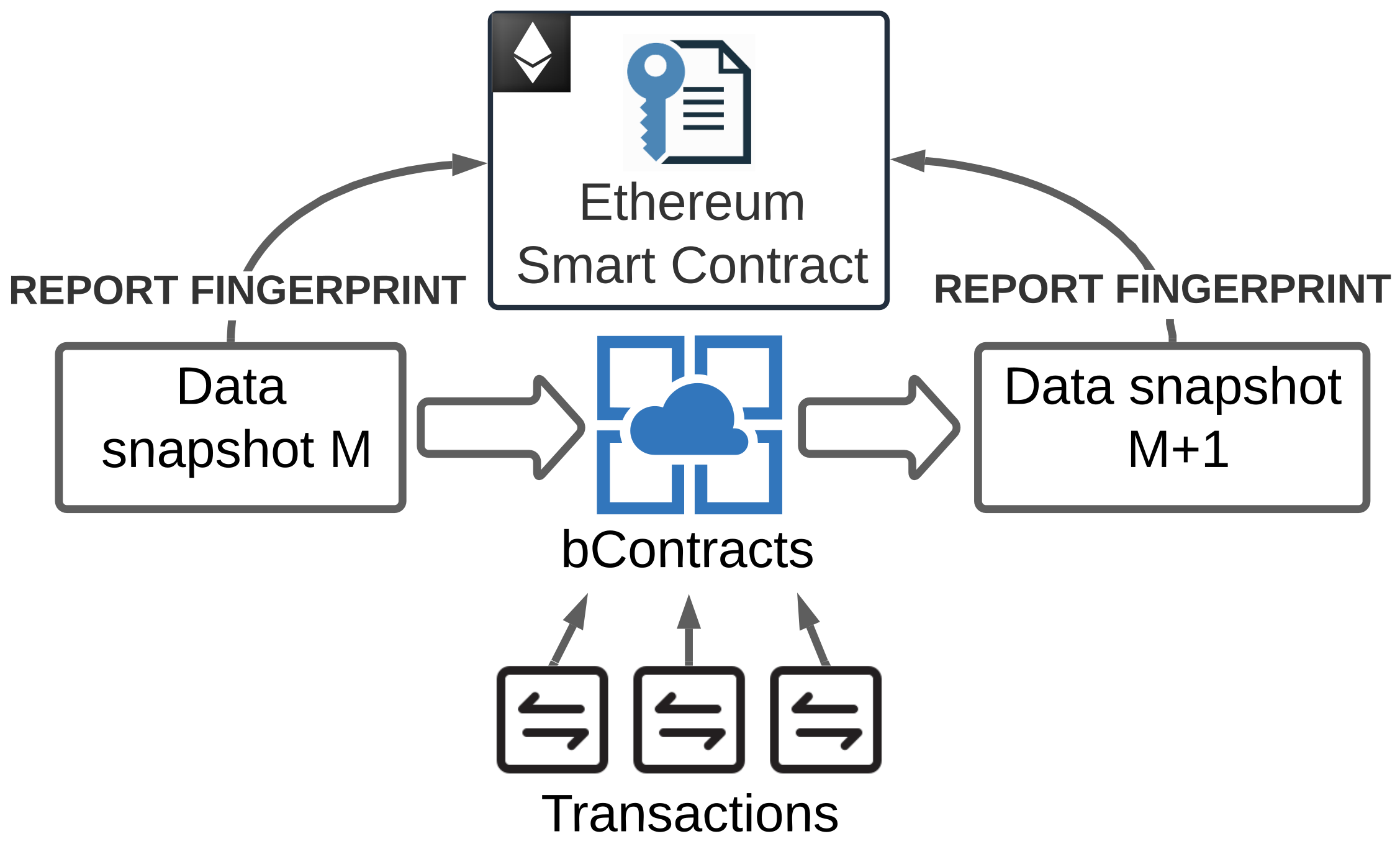}
    \caption{\textbf{\blockumulus state transition and data model.}
    }
    \label{fig:model}
    \vspace{-17pt}
\end{figure}

The code execution in \blockumulus is performed in decentralized \blockumulus smart contracts called bContracts, as shown in Fig.~\ref{fig:model}. The code of bContracts is openly accessible, so that the execution of transactions could be verified by anyone.
The functions of bContracts are invoked through signed transactions arriving at the network, and the code in bContracts can be executed by appropriate interpreters.  bContracts can be written in different programming languages, such as Python or JavaScript.

\subsubsection{\blockumulus Data Model}
All data in \blockumulus is openly accessible and managed via custom models implemented in the deployed bContracts. In order to store data as part of \blockumulus, each bContract must implement two interfaces: \emph{data fingerprinting} and \emph{data cloning}. Data fingerprinting is a function that produces a fingerprint of the bContract's current state or previously saved state. The data cloning function asks the contract to temporarily save its current state of data for subsequent fingerprinting. \blockumulus then combines all the fingerprints reported by the bContracts into a single hash called the \emph{data snapshot fingerprint}.

\subsubsection{Overlay Consensus}
\begin{figure}
    \centering
    \includegraphics[width=0.8\linewidth]{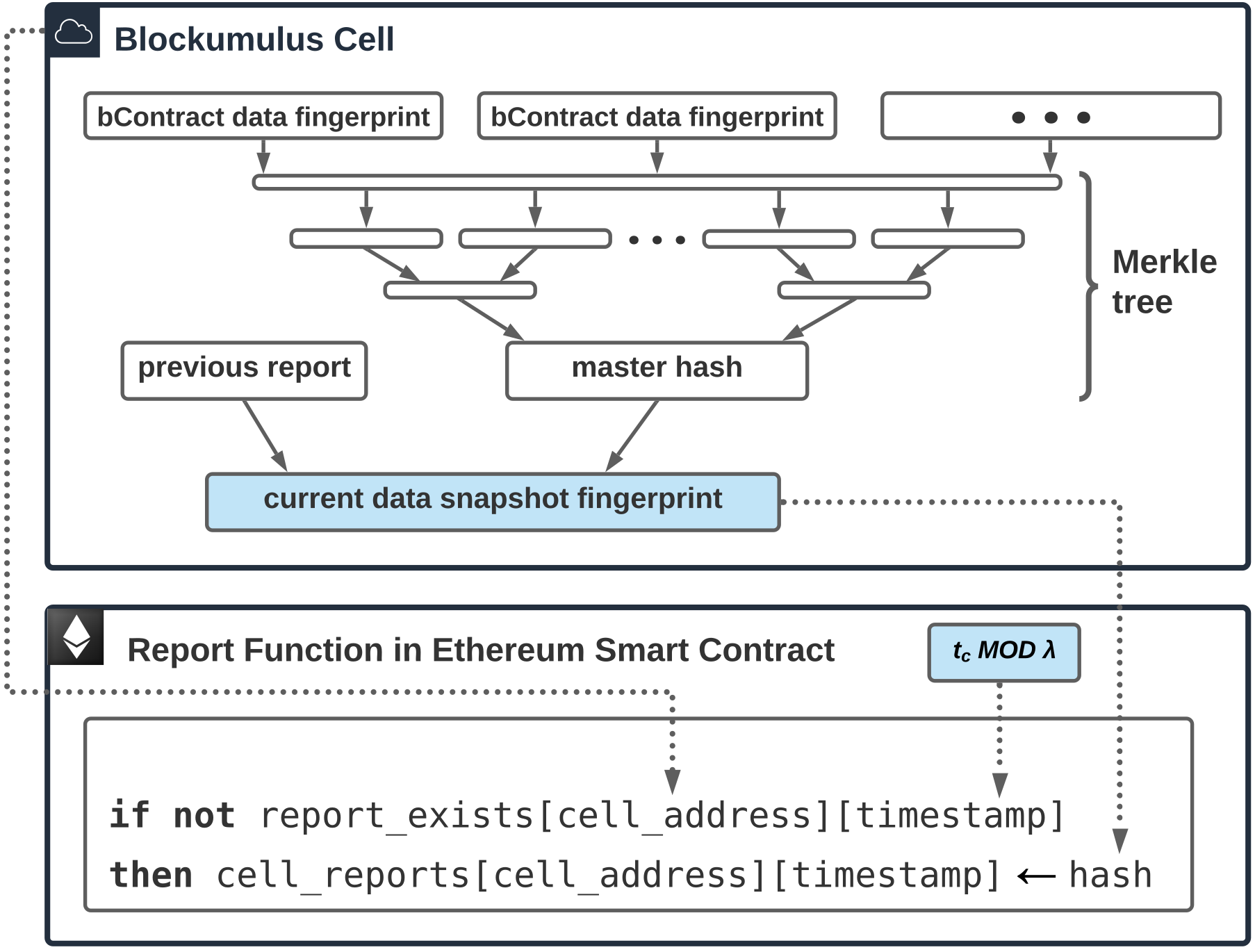}
    \caption{\textbf{Reporting of current cell state to the smart contract.}
    }
    \label{fig:report}
    \vspace{-10pt}
\end{figure}

The core idea of \blockumulus \emph{overlay consensus} is to periodically report the hashes of data snapshots to a dedicated smart contract deployed on a public blockchain, as shown in Fig.~\ref{fig:report}. Once the report is submitted, it cannot be altered. Subsequently, if the report does not match the publicly available and independently verifiable snapshot, the cell cannot be trusted. In essence, the Ethereum smart contract serves as an online barometer of liveness and integrity of the \blockumulus deployment. 

\blockumulus overlay consensus has two major differences with the traditional Nakamoto consensus observed in popular public blockchains. First, \blockumulus consensus uses correctness check instead of voting --- all incoming transactions are recorded, and there is only one correct way to execute them such that the existence of two conflicting transactions in different cells is ruled out (see Section~\ref{sec:transactions} for details). Second, all transactions are executed immediately, during the open session with the client, with a pre-defined decision deadline --- as a result, a consensus partitioning (called fork) is impossible in \blockumulus. Unlike in a distributed database, which stipulates identical query execution in all tables, \blockumulus provides autonomous but distinct execution environments for each individual bContract.  The contracts with mismatching fingerprints can be excluded from the consensus, and timely fingerprint reports can be guaranteed even if some contracts are unable to establish consensus within their respective contexts. The goal of each bContract is to assure that a transaction is executed identically across all the cells. To enforce this, after each transaction, the called bContract produces a fingerprint of its current data. If the fingerprints do not match, the bContract is temporarily excluded from the snapshot. As a result, each transaction entails an identical state transition of each cell in \blockumulus. If a cell becomes irresponsible or fails the verification, it is excluded from the consensus until the next report cycle.

\subsubsection{Report Timing}
Prior to deployment, the cloud consortium determines the system invariants that cannot be changed during the lifetime of the system. One of these invariants is the snapshot report period, denoted $\lambda$, which is measured in seconds. In \blockumulus, the report deadlines are all timestamps divisible by $\lambda$. Therefore, the last report deadline can be calculated as $t_{d} = t_{c} \text{~}MOD\text{~} \lambda$, where $t_{c}$ is the current timestamp. Thus, the upcoming report deadline is calculated as $t_{next} = \lambda + t_{c} \text{~}MOD\text{~} \lambda$. Every data snapshot, denoted $S_i$, has a serial number $i$, which is called \emph{the report cycle}, represented as $\tfrac{t_{d} - t_{0}}{\lambda}$, where $t_{0}$ is the deadline of the very first snapshot in the \blockumulus deployment. Subsequently, the \blockumulus protocol requires that each cell reports the snapshot $S_i$ by the end of cycle $i+1$ in order to be treated as valid during the cycle $i+2$.

\subsection{\blockumulus Components}
Next, we introduce the major components of \blockumulus: consortium of cloud cells, decentralized \blockumulus smart contracts (bContracts), clients, Ethereum smart contract, and independent auditors.

\subsubsection{Cloud Consortium}
The cloud consortium is a pre-defined set of \blockumulus cells. The number of cells should be sufficient to guarantee the availability of the system, but it should not be too large (i.e., 10 or less) to avoid performance degradation. Unlike peers in blockchain, multiple cells in \blockumulus are used to achieve the accessibility and fault-tolerance, rather than the consensus, which will be detailed in Sections~\ref{sec:incentive} and~\ref{sec:scalability}. Moreover, since clouds allow vertical scalability (i.e., adding resources to existing entities), a large number of cells (horizontal scalability) is not needed for performance advancement either. The size of the consortium and the set of identities of the participating cells are the invariants that must be decided at the time of deployment.

\subsubsection{\blockumulus Cell}
A \blockumulus cell is a network node on the cloud, which is sufficient for participating in \blockumulus consensus. A cell can be represented by a virtual machine, physical dedicated server, or a compute cluster --- whichever meets the demands of the system.

\subsubsection{bContracts}
\blockumulus smart contracts (bContracts), are decentralized programs deployed on \blockumulus, whose functionality is similar to smart contracts in Ethereum or chaincode in Hyperledger Fabric. There are two types of bContracts: system bContracts and community bContracts. The system bContracts are pre-deployed in \blockumulus, and they cannot be removed. The community bContracts are developed and deployed by clients.

\subsubsection{\blockumulus Clients}
\blockumulus client is a person or software that interacts with a deployed bContract. \blockumulus is a permissionless environment for clients, which means that clients do not have to register a \blockumulus account. However, akin to the ISP model for Internet access, a client should have a subscription to \blockumulus through one of the cells. The subscription, however, does not incur any control over the use of \blockumulus. The purpose of the subscription is to charge for data transferred or time period during which the subscription is active. This contrasts with the transaction fee collection observed in public blockchains. As a result, \blockumulus offers flexibility that allows cells to establish their own pricing policies to compete for customers.

\subsubsection{Ethereum Smart Contract}
Each \blockumulus deployment has a smart contract on Ethereum blockchain, which stores hashes of the reported snapshots. To avoid retrospective modification, the repeated reporting for the same timestamp is prohibited by the logic of the smart contract.

\subsubsection{\blockumulus Auditors}
\begin{figure}
    \centering
    \includegraphics[width=0.8\linewidth]{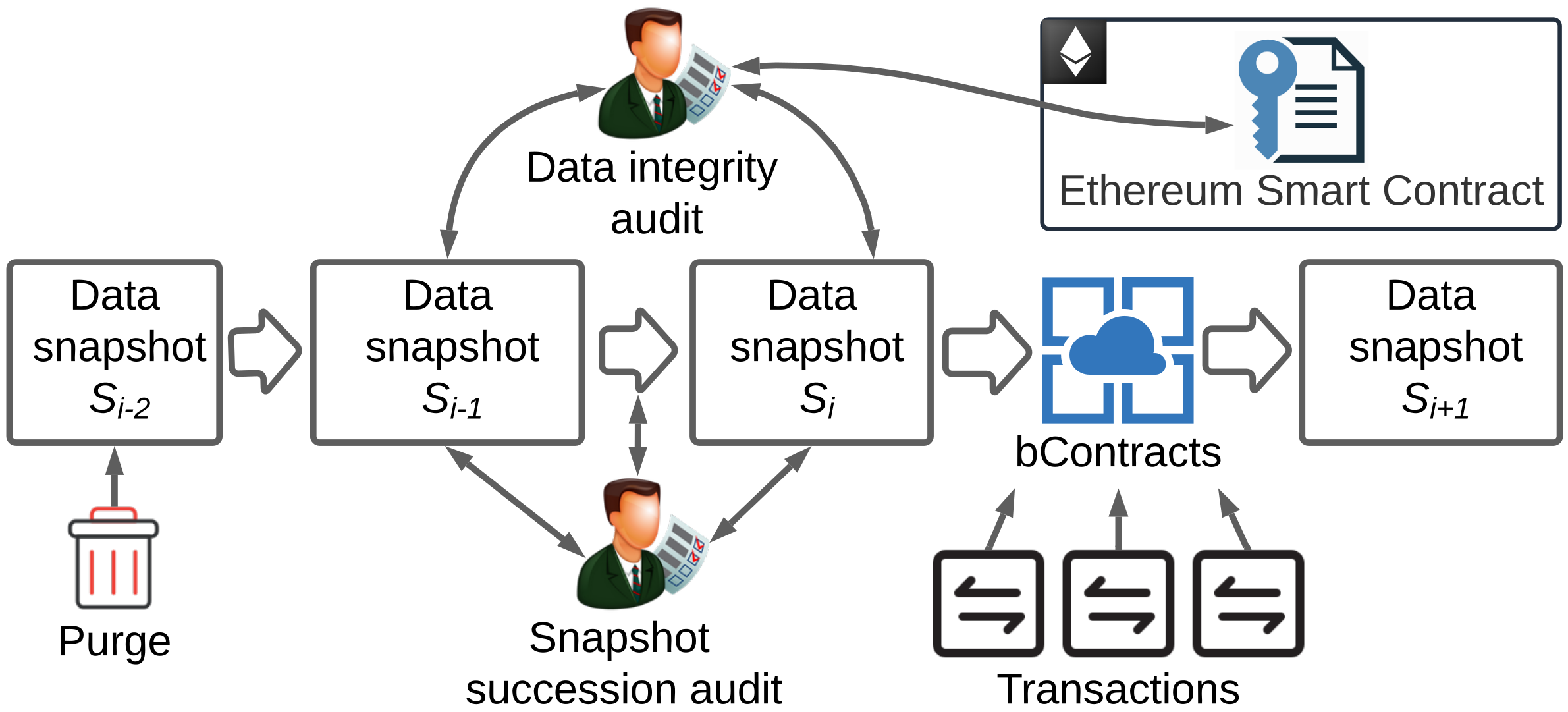}
    \caption{\textbf{\blockumulus audit procedure.}
    }
    \label{fig:audit}
\end{figure}

Akin to public blockchain, \blockumulus is an open-data system with transparent execution, i.e., \blockumulus data is available to everyone, and everyone can independently trace state transition between a given pair of subsequent data snapshots. Auditors are voluntary permissionless participants that run software to oversee the integrity of the \blockumulus deployment. The community auditing model, which demonstrated its efficiency in public blockchains, is also employed in \blockumulus. Auditors can be a paid participants, community enthusiasts, security bounty hunters, or academic researchers. Moreover, cells in the consortium can perform cross-audit. The process of auditing requires only a server and the auditing software that is running on this server to monitor the integrity of \blockumulus. Fig.~\ref{fig:audit} shows the procedure of the \blockumulus audit. The auditing software performs two major tasks: snapshot succession audit and data integrity audit. The snapshot succession audit is the verification that all the transactions processed by all bContracts between two reports indeed entail a state transition from one data snapshot into another. The data integrity check verifies that: a) the snapshot fingerprints have been reported to the smart contract on time; and b) the fingerprints in reports match the actual data in the cells.

\subsection{\blockumulus Cell Architecture}

\begin{figure}
    \centering
    \includegraphics[width=0.8\linewidth]{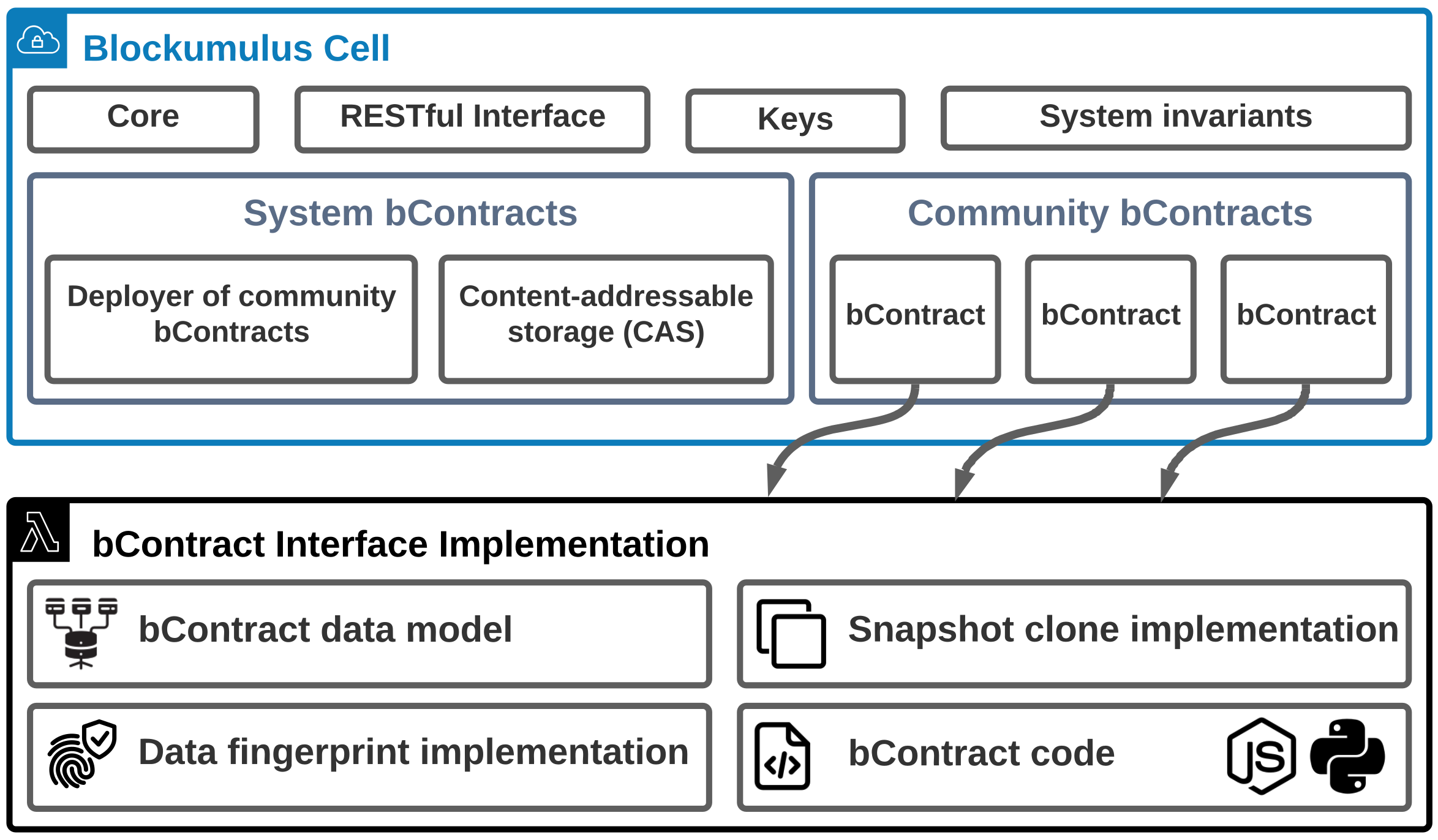}
    \caption{\textbf{\blockumulus components and bContracts.}
    }
    \label{fig:cell}
    \vspace{-15pt}
\end{figure}

In this section, we take a closer look at the architecture of a cell, which is shown in Fig.~\ref{fig:cell}.

\subsubsection{\blockumulus Core}
\blockumulus cell Core is responsible for networking, cryptography, synchronization, protocol, process and thread management, signature and authenticity verification, transaction parsing, data encoding and decoding, and communication with the smart contract.

\subsubsection{Uniform RESTful Interface}\label{sec:restful}
\blockumulus assumes six vectors of communication: client-cell, cell-cell, auditor-cell, cell-blockchain, auditor-blockchain, and client-auditor. The client-cell, cell-cell, and auditor-cell communications have a uniform RESTful interface. Specifically, each request is either GET or POST HTTP request with the body formally represented as the set $\mathcal{M} = \{\mathcal{P}= \langle A_{s}, A_{r}, O, \eta, \tau, t, \mathcal{D} \rangle, Sig_{s}(\mathcal{P}) \}$, where $\mathcal{P}$ is the payload of the message, and $Sig_{s}$ is the ECDSA signature calculated via the private key of the sender. The tuple $\mathcal{P}$ has the following components: $A_s$ is the public address of the sender, $A_r$ is the public address of the intended recipient, $O$ is the operation code, $\eta$ is a random nonce used as a message ID, $\tau$ is the ID of the message that $\mathcal{M}$ is replying to (if applicable), $t$ is the current timestamp, and $\mathcal{D}$ is the data, whose format is determined by $O$.

\subsubsection{Keys}
Each cell uses an Ethereum account to represent itself within \blockumulus. The set of public addresses\footnote{In Ethereum, a public address of an account is the 160-bit prefix of the Keccak256 hash of the account's public key.} of \blockumulus cells is fixed for each deployment and is hard-coded in the Ethereum smart contract.

\subsubsection{System Invariants}
Some parameters of a \blockumulus deployment that remain constant for a lifetime are called the \emph{system invariants}. Examples of system invariants are: unique deployment ID, identities of the cells, reporting period $\lambda$, initial timestamp $t_0$, etc. However, the IP addresses of cells are not among the invariants, which allows cells to change location, or network configuration --- we assume that these settings are exchanged between cells.

\subsubsection{System bContracts}
The system bContracts are pre-implemented as part of \blockumulus, and they cannot be removed. These bContracts deliver essential functionality to the system, and their number can grow as \blockumulus framework evolves. The current version of \blockumulus includes two system bContracts: \emph{community bContract deployer}, and \emph{content-addressable storage (CAS)}. The community app deployer serves as an interface for developers to add their community bContracts to \blockumulus. The CAS contract has two major functions: a) it allows to store large files outside of data models of community bContracts, thereby significantly improving the performance of fingerprinting and cloning; and b) it establishes a secure communication channel between bContracts, which are otherwise autonomous and isolated.

\subsubsection{Community bContracts}
Community bContracts are developed and deployed by users of \blockumulus. The cells have no power to modify, censor, or control these contracts. The deployer of a community bContract can specify the ownership and other parameters of the contract, including the ability to destroy one.

\subsubsection{bContract Interface}
In order to create a bContract, the developer should implement a standard bContract interface, which includes smart contract data model, data fingerprinting, and snapshot cloning. Then, the developer writes the bContract code for the interpreter specified in the configuration.

\subsection{\blockumulus Protocol}

\subsubsection{Data Snapshots and Fingerprinting}
\blockumulus data is stored in bContracts according to their respective data models. For example, one bContract can store data in binary files, while others may use SQLite. To prevent operations with large data instances, bContracts can upload data blobs to \blockumulus CAS, and refer to these blobs via their hashes. \blockumulus performs CAS reference counting, purging CAS entries only when their reference counters reach zero.

\subsubsection{Operation Lifecycle}

\begin{figure}
    \centering
    \includegraphics[width=\linewidth]{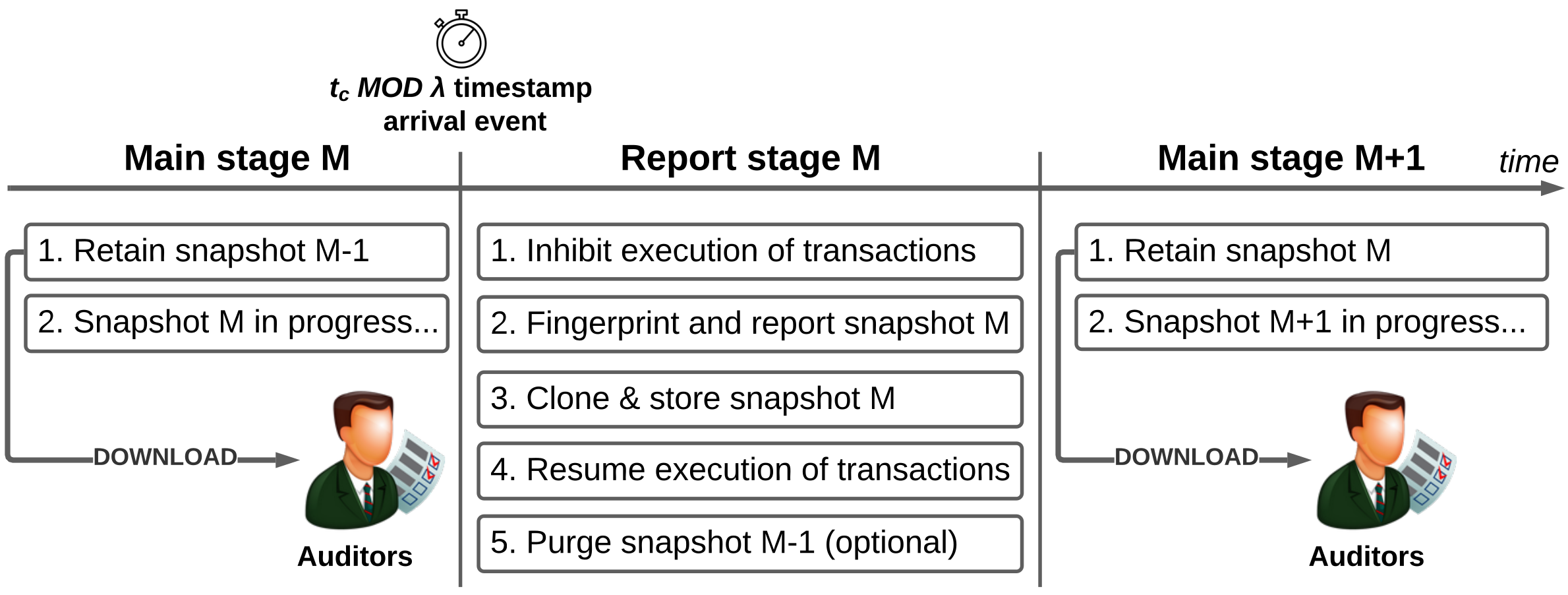}
    \caption{\textbf{\blockumulus lifecycle.}
    }
    \label{fig:stages}
    \vspace{-15pt}
\end{figure}

Fig.~\ref{fig:stages} shows the lifecycle of \blockumulus involving an oscillation of two stages: \emph{main stage} and \emph{report stage}. In the main stage, which is longer than the report stage, \blockumulus actively accepts and processes incoming transactions that shape the current data snapshot. During the main stage, auditors download the previous data snapshot for review and storage. In the report stage, \blockumulus accepts transactions, but instead of executing them, it queues them in a buffer. Once the current snapshot is fingerprinted, \blockumulus continues executing incoming and queued transactions. Also, as soon as the fingerprint is ready, the cell saves it in the smart contract. However, at this point, the execution of the incoming transactions resumes because the execution inhibition is needed only for calculating the fingerprint, not for smart contract submission.

\subsubsection{Transactions}\label{sec:transactions}

\begin{figure}
    \centering
    \includegraphics[width=0.9\linewidth]{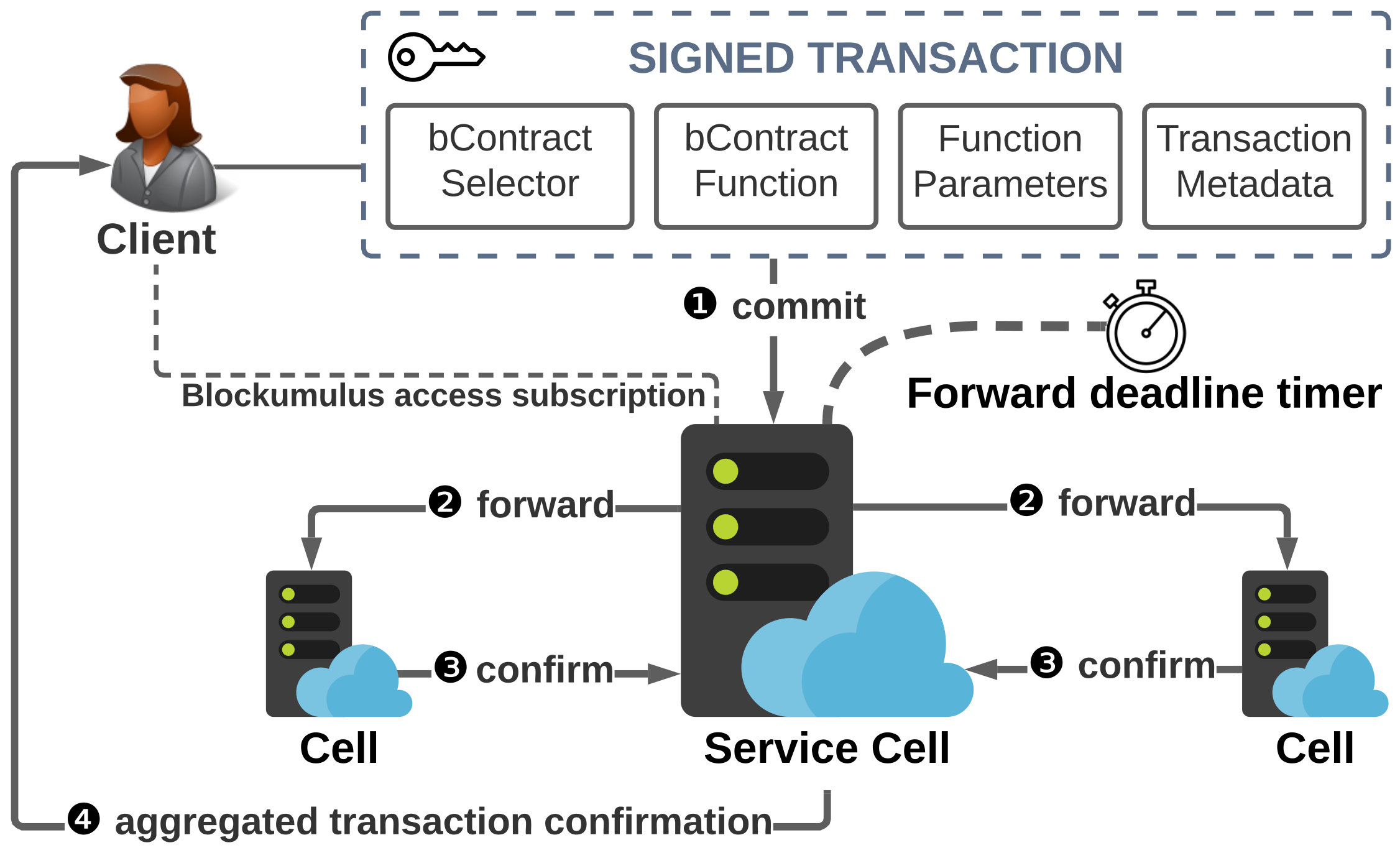}
    \caption{\textbf{\blockumulus transaction workflow.} \ding{182}: Client creates a transaction and commits it to the the blockumulus cell with which they have a \blockumulus access subscription; \ding{183}: the service cell verifies the authenticity of the transaction, and forwards it to all the other cells in the consortium; \ding{184}: the cells of the consortium process the transaction and send a signed confirmation back to the service cell within a strict deadline; \ding{185}: the service cell executes the transaction, serializes the confirmations into an aggregated receipt, and sends it to the client as a reply to the initial commit request.}
    \label{fig:transaction}
    \vspace{-10pt}
\end{figure}

Fig.~\ref{fig:transaction} shows a general overview of a \blockumulus transaction. The transaction begins with a client creating a transaction message $\mathcal{M}$, which is signed and sent to the the \blockumulus cell, called the \emph{service cell}, with which the client has an access subscription. The service cell first authenticates the transaction by confirming that the transaction message is signed by the user with the same identity (public address) as the one found in the transaction message. Then, the service cell forwards the transaction to all the cells in the consortium. After that, the cells of the consortium verify and execute the transaction and send a signed confirmation back to the service cell within a pre-determined short time frame. If the forwarded transaction is not processed by all cells until the established deadline, the transaction reverts. If a cell misses the deadline more often than a pre-determined threshold, it is temporarily excluded from the consensus upon mutual agreement with the other cells. Finally, the service cell verifies the fingerprints of the resulting data snapshots reported by the other cells, and executes the transaction by itself. If the result of the execution matches the fingerprints reported by the other cells, the service cell serializes the confirmations into an aggregated receipt, and sends it to the client as a reply to the initial commit request, which constitutes the transaction confirmation event with a multi-signature cryptographic proof (in the format described in Section~\ref{sec:restful}).

\subsubsection{Incentive for Cooperation}~\label{sec:incentive} Here, the incentive for cooperation is discussed through the P2P network perspective. 
Unlike in public blockchain consensus (e.g., Nakamoto consensus), \blockumulus is designed in a way to encourage cooperation and make cheating unbeneficial. The combination of synchronous execution, fixed cell topology, open data, transparent execution, and payment model separated from consensus create an arrangement in which cells have no incentive to cheat. Moreover, each cell benefits from fast and successful execution of transactions by all other cells in the system. The following theorem confirms that competition for voting power, typical for blockchains, is not pertinent to \blockumulus.

\vspace{5pt}
\noindent\textbf{Theorem 1:} \emph{The minimum required number of valid cells in \blockumulus overlay consensus is the same for all $M \geq 2$.}

\noindent\textbf{Proof:}
As per design of \blockumulus, the auditor software verifies that the deployment has at least one cell $i$ that maintains the succession of reported snapshots $S_{i,j}$ and correctness of the corresponding smart contract reports $R_{i,j}$, i.e.:
\begin{multline}
    \exists \: 1 \leq i \leq M \; \forall \text{~} 1 \leq j \leq \tfrac{{t_{c} \text{~}MOD\text{~} \lambda - t_{0}}}{\lambda} : \\
    S_{i,j} \xrightarrow[]{succession} S_{i, j+1} \text{~}\wedge\text{~} H(S_{i,j}) = R_{i, j},
\end{multline}
where $H$ is the hash function used for fingerprinting in \blockumulus. Suppose that $M=2$, one cell is valid, while all other cells may or may not be compromised or cheating. In this case, formula (1) evaluates to ``true", because either Cell 1 is valid, or Cell 2 is valid, or both of them are valid. Now, suppose that $M=Q$ ($Q>2$), and one cell is valid, while all other cells may or may not be compromised or cheating. In this case, formula (1) again evaluates to ``true",  because there is a cell with an index in the range $[1,Q]$, which maintains succession of snapshots and correctness of the fingerprint reports. Therefore, the minimal number of cells required for the overlay consensus is always 1.
$\blacksquare$

\section{Scalability Analysis}\label{sec:scalability}
In this section, we formally explore the scalability of \blockumulus through an asymptotic complexity analysis. All the assumptions in this section follow the real implementation of the system described later in Section~\ref{section:evaluation}.

%\subsection{Notation and Assumptions}
Here, we assume that $K$ clients submit $N$ successful transactions to a \blockumulus deployment with $M$ cells. We use the symbol $c$ to denote a constant value that does not grow as the system scales.

\noindent\textbf{Number of Cells.}
Unlike blockchain, in which an increase of the number of nodes benefits decentralization, the \blockumulus overlay consensus requires only one valid cell to sustain normal operation, including prevention of conflicting transactions, such as double spending. As per \emph{Theorem~1}, proven in Section~\ref{section:systemdesign}, adding more cells does not enhance the decentralization of a \blockumulus deployment. Thus, we neither require the number of cells $M$ to be scalable, nor do we assume its scalability. The two reasons for using multiple cells in \blockumulus is to enhance availability of the system through replication and to increase the diversity of \blockumulus access providers.

\subsection{Transaction Latency}
Transaction latency is the total delay experienced by the client between the initiation of a transaction until the confirmation of its completion. The cumulative transaction delay in the system, denoted $L_{delay}$, can be expressed as $L_{delay} = N \cdot (D_{1} + \max_{i = 2}^{M}{(D_i + D_i^*)} + D_c)$, where $D_1$ is the delay of sending a transaction to the service cell, $D_i$ is the delay in forwarding the transaction to cell $i$, $D_i^*$ is the delay of response from cell $i$ to the service cell, and $D_c$ is the delay of sending the response to the client. We also assume that $D_i + D_i^* < \delta$ for all $i > 1$, where $\delta$ is maximum transaction forwarding delay. Each of the $N$ transactions begins with the client sending it to the service node, which simultaneously forwards the transaction to all the other cells, followed by an immediate parallel response from these cells to the service cell. Then, it finishes by sending the aggregate response to the client from the service cell. Now, since $D_1$, $\delta$, and $M$ do not grow with increased number of transactions, the transaction latency complexity can be presented as $L_{delay} = N \cdot (c + c \cdot c + c) = \mathcal{O}(N)$. Therefore, the transaction latency in \blockumulus grows linearly with the number of transactions. Section~\ref{sec:txn-latency} further shows that the transaction latency remains low even when the cells are deployed on low-tier cloud servers with an extreme transaction load.

\subsection{Communication Overhead}\label{sec:comm-overhead}
Transaction communication overhead is the cumulative amount of data transferred within \blockumulus in the course of $N$ transactions. The communication overhead $L_{data}$ of $N$ \blockumulus transactions can be expressed as follows:
\begin{multline}\label{eq:ldata}
    L_{data} = N \cdot [H_{c} + P_{c} + (M-1) \cdot (H_{1} + H_{c} + P_{c}) \\
    + \sum_{i = 2}^{M}{(H_{i} + P_{i})} + \sum_{i = 1}^{M}{(H_i + P_i)]},
\end{multline}
where $H_c$ is the header sent by a client, $P_c$ is the payload sent by a client, $H_i$ is the header sent by a cell $i$, and $P_i$  is the payload sent by the cell $i$. Since headers and payloads of messages do not become bigger with more transactions, and the number of cells remains constant, Eq.~(\ref{eq:ldata}) can be reduced as $L_{data} = N \cdot [2c + c \cdot 3c + c \cdot 2c + c \cdot 2c] = \mathcal{O}(N)$. Therefore, as the number of transactions grows, the communication overhead also experiences a linear increase. In Section~\ref{sec:comm-overhead} we show that this complexity is practically amenable and does not lead to bottlenecks even under an extreme transaction load.

\subsection{Data Storage}

We assume that each transaction in \blockumulus leaves a data footprint $U_i$, which is replicated across participating cells, and also appears in three snapshots\footnote{\blockumulus uses the CAS subsystem to prevent unnecessary replication of the same data across several snapshots. However, since our analysis pursues the upper bound complexity, we assume 100\% replication of the data.}: the snapshot currently being built, and also two previous snapshots left for auditing. The data storage can be written as $L_{storage} = 3 \cdot M \cdot \sum_{i = 1}^{N}{U_i}$. Since the number of cells $M$ and each of the size of stored data items $U_i$ do not grow with the increasing number of transactions and users, the following reduction takes place: $L_{storage} = 3 \cdot c \cdot N \cdot c = \mathcal{O}(N)$. Therefore, the complexity of the stored data is linear with respect to the number of transactions.

\subsection{Computation}
In our \blockumulus compute analysis we take into consideration the processing performed both by cells and by auditors. We further assume that the number of auditors is linearly proportional to the number of users $K$, i.e., certain percentage of users serve as auditors. Then, the cumulative computation overhead can be represented as $L_{compute} = K \cdot \sum_{i = 1}^{N}{(C_i)} + M \cdot \sum_{i = 1}^{N}{(C_i)}$, where $C_i$ is the amount of computation required for processing a single transaction $i$ on a single computer. Since each computational load and the number of cells remain the same with growing number of transactions and users, we perform the following reduction: $L_{compute} = K \cdot N \cdot c + c \cdot N \cdot c = \mathcal{O}(KN)$. Therefore, the compute overhead of \blockumulus has a linear dependency upon both the number of users and the number of transactions, which suggests that the cells may require to proportionally increase their compute power as the number of users grows. Since users are expected to pay for \blockumulus access, the above requirement is unlikely to form a scalability bottleneck.

\subsection{Snapshot Reporting}

Each \blockumulus cell reports fingerprints to the smart contract with constant frequency $F = \tfrac{1}{\lambda}$. By representing the report timeline through $R$, the blockchain fee overhead is as follows: $L_{fee} = M \cdot R \cdot F$. Since the number of cells $M$ is fixed, the fee does not change over time, and the report frequency is also fixed, i.e., $L_{fee} = c \cdot c \cdot c = \mathcal{O}(1)$. Therefore, as a \blockumulus deployment grows, the fee overhead remains in the same order.

\vspace{-5pt}
\section{Security Analysis}
Blockchain is a target of a wide range of security threats, from consensus-based attacks~\cite{nakamoto2019bitcoin} to social engineering attacks~\cite{ivanov2021targeting}, and \blockumulus is not an exception. In this section, we scrutinize critical scenarios that pose security threats to a \blockumulus deployment, and we show how \blockumulus addresses these challenges.

\subsection{Double Spending}
A double spending is a situation in which two mutually exclusive transactions are executed by a distributed system, such as repeated transfer of the same cryptocurrency balance. Consider a situation in which Alice, who has 10 \emph{crypto} coins, creates a transaction that sends 10 coins to Bob, and another transaction with identical timestamp that sends 10 coins to Charlie. After that, Alice simultaneously submits one of these transactions to \blockumulus through Cell 1, and another one through Cell 2. Assume that the transaction storage of \blockumulus is properly implemented with a mutex-based storage (i.e., the one that does not permit simultaneous writing operations), which can be achieved through file locks or ACID databases. The two transactions will be saved in the ledger in the order of their arrival. Subsequently, the transaction that is executed second will be rejected, effectively preventing the double spending. Furthermore, \blockumulus transactions are executed synchronously by all cells. Unlike blockchain, which allows a temporary partition into peers that have already processed a transaction and peers that have not, \blockumulus prohibits temporary asynchrony using the synchronous execution with a mutex-based storage. Therefore, the situation where Bob received 10 coins from Alice according to one cell and Charlie received 10 coins according to another cell is impossible.

\subsection{Transactions Filtering Attack}\label{sec:filtering}
\blockumulus cells might prevent routing of a certain transaction to a bContract via a transactions filtering attack. For example, consider a bContract that re-invests dividends if an investor fails to withdraw them until a certain deadline. The invested business might bribe the cloud consortium to filter out the withdrawal transaction --- in which case the auditors will not be able to detect any anomaly. \emph{In \blockumulus, we address this issue by  enforcing the execution of a transaction via the Ethereum smart contract.} If a transaction is censored, it can be submitted directly to the smart contract, and the system protocol stipulates the necessity to execute all transactions submitted in this way. Since the smart contract is not under any party's control, users have the ability to enforce a transaction even when \blockumulus has only one operational cell.

\subsection{Consortium Conspiracy}
The cells might conspire to tamper with the snapshots in three possible ways: 1) by modifying an existing transaction, 2) by removing an existing transaction, or 3) by injecting a new transaction. If an existing transaction is modified, it will immediately break the verification of the transaction signature generated by the sender. If an existing transaction is removed before the report is submitted to the smart contract, the receipt of this transaction signed by the cell becomes the proof of malice by the cell. Finally, if a new transaction is added before the report, it is a legitimate way to change data in the snapshot and does not need to be defended against. Another type of consortium conspiracy is a system-wide subscription ban of a user by all \blockumulus providers. Fortunately, this type of conspiracy can be easily prevented in the same way as in the case of transaction filtering attack (see Section~\ref{sec:filtering}), i.e., by letting users submit contingency transactions to the Ethereum smart contract.

\subsection{Compromised Cells}
An attacker might compromise one or several Blockumulus cells to skew the overlay consensus. Let us consider the worst-case scenario, in which the attacker gained full access to the majority of cells in a Blockumulus deployment to cause the Byzantine Fault event. In this case, a consensus node cannot verify the true state of the system based on the testimonies from the other nodes. However, \blockumulus is not prone to the Byzantine Fault scenario, because the Ethereum smart contract, deployed on a Byzantine Fault Tolerant (BFT) blockchain, prevents the cells from delivering inconsistent testimonies to different parties.

\section{Implementation and Evaluation}\label{section:evaluation}

We implement \blockumulus framework and evaluate its transaction latency, communication overhead, transaction throughput, and operation cost. To account for different configurations, we test the system performance with three different sizes of the cloud consortia: $N=2$, $N=4$, and $N=8$.

\subsection{Implementation}
We implement the full stack of \blockumulus for evaluation and proof of concept.
%, and in the spirit of open research we made its source code available at \url{https://github.com/nick-ivanov/blockumulus}. 
The \blockumulus API is implemented using Web3.js 1.3.0, and node-rest-client 1.3.1. The \blockumulus core framework is implemented using Node.js 10, Express 4.17, and Web3.js 1.3. We deploy 8 \blockumulus cells on individual Ubuntu 20.04 servers on Microsoft Azure cloud. Then, we implement the Ethereum smart contract using Solidity 0.8.0, with the test deployment available on Ropsten network at \texttt{\small0x2F2980067A524a9A12C46354D62B8D769Ee119AB}. The implementation includes 2,553 lines of code. To demonstrate the performance of \blockumulus, we implement a sample bContract called \emph{FastMoney} using Python 3.6 and Web3.py 5.13 (for fingerprinting), which delivers a decentralized digital currency. Then, we implement the user clients for \emph{FastMoney} and CAS in JavaScript and Web3.js, which are used for automated evaluation, as described below.

\begin{figure*}
\captionsetup[subfigure]{justification=centering}
\centering
\begin{subfigure}{.32\textwidth}
  \centering
  \includegraphics[height=1.5in]{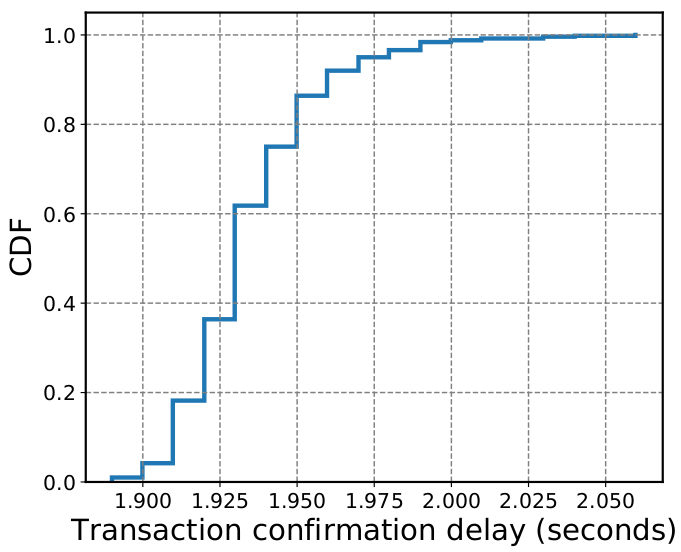}
  \caption{2 cells}
  \label{fig:latency-2cells}
\end{subfigure}%
\begin{subfigure}{.32\textwidth}
  \centering
  \includegraphics[height=1.5in]{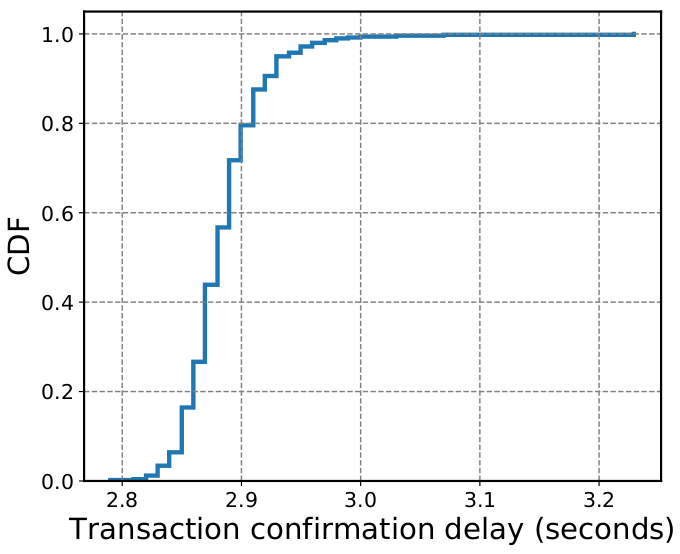}
  \caption{4 cells}
  \label{fig:latency-4cells}
\end{subfigure}%
\begin{subfigure}{.32\textwidth}
  \centering
  \includegraphics[height=1.5in]{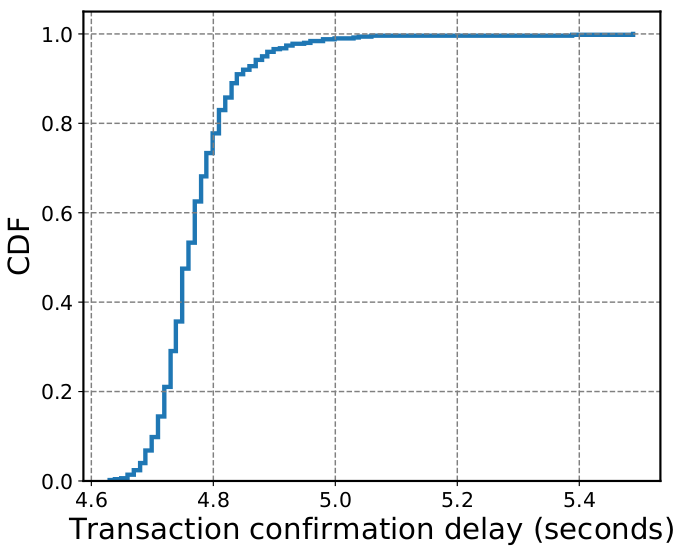}
  \caption{8 cells}
  \label{fig:latency-8cells}
\end{subfigure}

\caption{\textbf{Transaction latency for \emph{FastMoney} funds transfer with different sizes of cloud consortia based on 500 requests.}
}
\label{fig:latency}
\vspace{-10pt}
\end{figure*}

\subsection{Test Setup}

\noindent\textbf{System Under Test.}
Our system under test (SUT) includes a set of cell deployments and an Ethereum smart contract on the Ropsten testnet. For latency evaluation, we use \blockumulus cell deployments with three different sizes of cloud consortia: $N=2$, $N=4$, and $N=8$. For each cell, we deploy an Azure B1ms instance with Ubuntu 20.04 LTS.

\noindent\textbf{Test Harness.}
We use Blockumulus API to create custom test clients with the additional functionality of generating a random account for each request to simulate \emph{different} clients and avoid potential caching of data related to a single account. Then, we deploy 8 \emph{client pools}, which are Azure Virtual Machines running Ubuntu 20.04 LTS each, scattered across different geographic regions for better simulation of a real-world distribution of clients.

\noindent\textbf{Evaluation Metrics.}
We evaluate transaction latency, communication overhead, transaction throughput, and operational cost of our prototype.

\subsection{Transaction Latency}\label{sec:txn-latency}
We evaluate the transaction latency of our \blockumulus deployment by measuring the time between submitting a transaction until the acquisition of the receipt. We conduct two latency evaluation tests: distribution of delays of standalone transactions under normal load, and transaction latency under the load of a large number of simultaneous transactions.

The results of the first experiment are shown in Fig.~\ref{fig:latency}. In this experiment, we measure transaction latency for the funds transfer in \emph{FastMoney} bContract with the sizes of the cloud consortia of 2, 4, and 8 cells. For each consortium size, we execute 500 consecutive transactions and measure their confirmation delays. When the size of consortium is 2, 90\% of transactions execute in under 2 seconds. When we double the size of consortium, the upper boundary of the confirmation delay of 90\% of transactions increases by around 50\%, which is slower than the increase of the number of cells. By doubling the number of cells again up to 8, we observe again that 90\% of transactions finished in under 5 seconds, which is around 66\% greater than in the case of 4 cells. Thus, the result is indicative that the growth of the transaction latency is slower than the number of cells.

In the second transaction latency measurement, we conduct a stress test with multiple transactions issued at the same time. For this experiment, we use the CAS system bContract, and run 9 experiments: with 5,000, 10,000, and 20,000 transactions, for each of the consortia sizes seen in the previous experiment, i.e., 2 cells, 4 cells, and 8 cells. Similar to the previous experiment, we can observe that in each configuration, as the number of transactions doubles, the transaction confirmation time increases by a lesser factor.

\begin{figure}
\captionsetup[subfigure]{justification=centering}
\centering
\begin{subfigure}{.3\linewidth}
  \centering
  \includegraphics[height=2.4in]{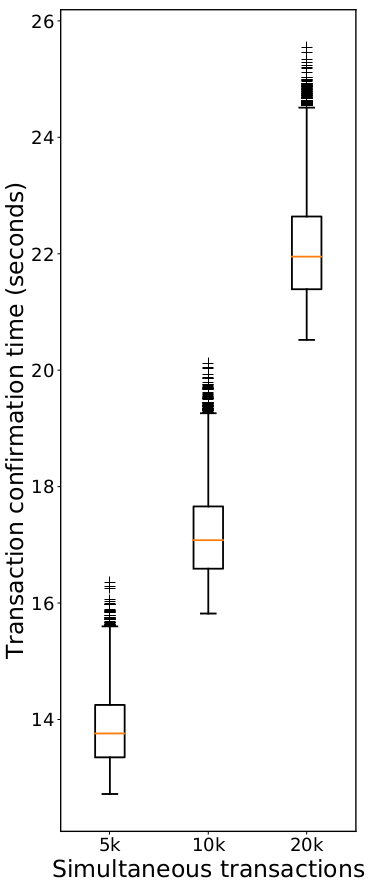}
  \caption{2 cells}
  \label{fig:latency-2cells}
\end{subfigure}%
\begin{subfigure}{.3\linewidth}
  \centering
  \includegraphics[height=2.4in]{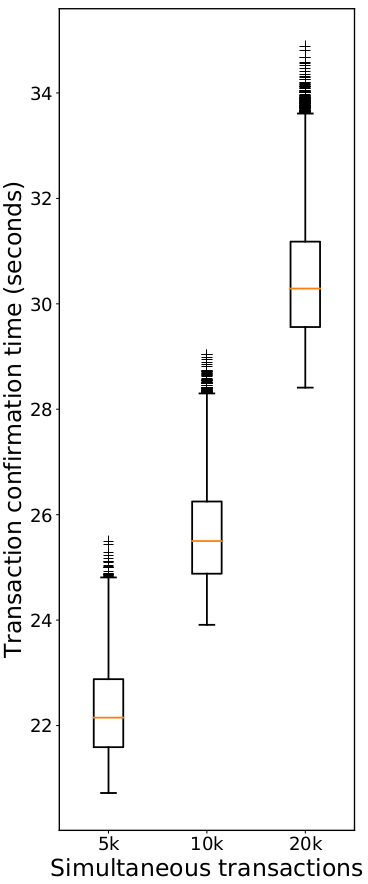}
  \caption{4 cells}
  \label{fig:latency-4cells}
\end{subfigure}%
\begin{subfigure}{.3\linewidth}
  \centering
  \includegraphics[height=2.4in]{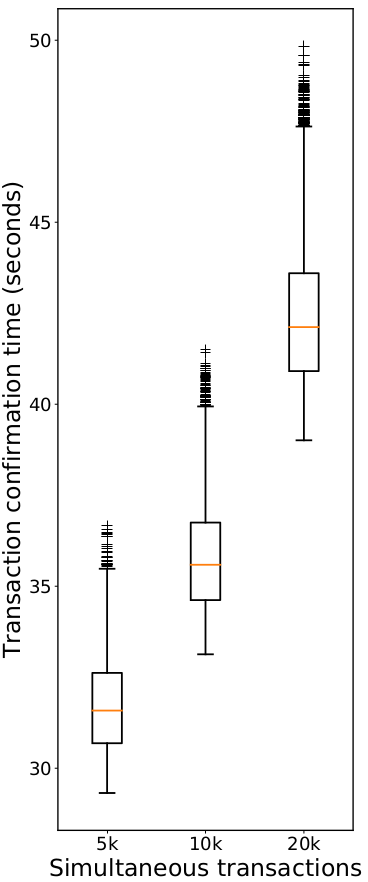}
  \caption{8 cells}
  \label{fig:latency-8cells}
\end{subfigure}

\caption{\textbf{Transaction latency for simultaneous CAS upload requests with different sizes of cloud consortia.}}
\vspace{-10pt}
\label{fig:latency-multi}
\end{figure}

\subsection{Communication Overhead}

Table~\ref{tab:overhead} shows the TCP overhead observed in \blockumulus while processing a transfer transaction with \emph{FastMoney} bContract. In order to observe the communication between cells, we create a 2-cell \blockumulus deployment on a local machine and run WireShark, in which we use the \emph{Follow TCP Stream} function to observe the cumulative traffic of each communication for each direction. The results shows that, in the worst case, the largest communication is around 4 Kbytes per transaction in downlink direction. A speed test using the Ookla software on several Azure servers revealed the available bandwidth around 8.5 Gbps in the downlink direction and around 1 Gbps in the uplink direction. Since the overhead of a \emph{FastMoney} transaction does not exceed 4 Kbyte, the 1 Gbps server bandwidth is capable to transfer the data of more than 30,000 transactions per second, which exceeds the average throughput of all credit card transactions in the world~\cite{creditcard}.

\begin{table}[]
    \centering
    \caption{\textbf{Communication overhead in FastMoney (bytes).}}
    \label{tab:overhead}
    \begin{tabular}{|c|c|c|c|c|c|c|}
        \hline
         \multirow{2}{*}{\textbf{Communication$^{\mathrm{a}}$}} & \multicolumn{2}{c|}{\textbf{2 cells}} & \multicolumn{2}{c|}{\textbf{4 cells}} & \multicolumn{2}{c|}{\textbf{8 cells}}  \\
         \cline{2-7}
         & \textbf{\textit{in}} & \textbf{\textit{out}} & \textbf{\textit{in}} & \textbf{\textit{out}} & \textbf{\textit{in}} & \textbf{\textit{out}} \\
         \hline\hline
         $CL \leftrightarrow C$: fingerprint & \multirow{1}{*}{1,200} & \multirow{1}{*}{516} & \multirow{1}{*}{2,179} & \multirow{1}{*}{516} & \multirow{1}{*}{4,135} & \multirow{1}{*}{520}  \\
         \hline
         $CL \leftrightarrow C$: payment & \multirow{1}{*}{1,140} & \multirow{1}{*}{559} & \multirow{1}{*}{2,059} & \multirow{1}{*}{559} & \multirow{1}{*}{3,895} & \multirow{1}{*}{563} \\
         \hline
         $CL \leftrightarrow C$: forward & \multirow{1}{*}{667} & \multirow{1}{*}{947} & \multirow{1}{*}{667} & \multirow{1}{*}{946} & \multirow{1}{*}{667} & \multirow{1}{*}{947} \\
         \hline
    
    \multicolumn{7}{l}{$^{\mathrm{a}}$ $CL \leftrightarrow C$: between client and cell; $C \leftrightarrow C$: between two cells.}
    
    \end{tabular}
\vspace{-15pt}
\end{table}

\subsection{Transaction Throughput}
\begin{figure}
    \centering
    \includegraphics[width=2.4in]{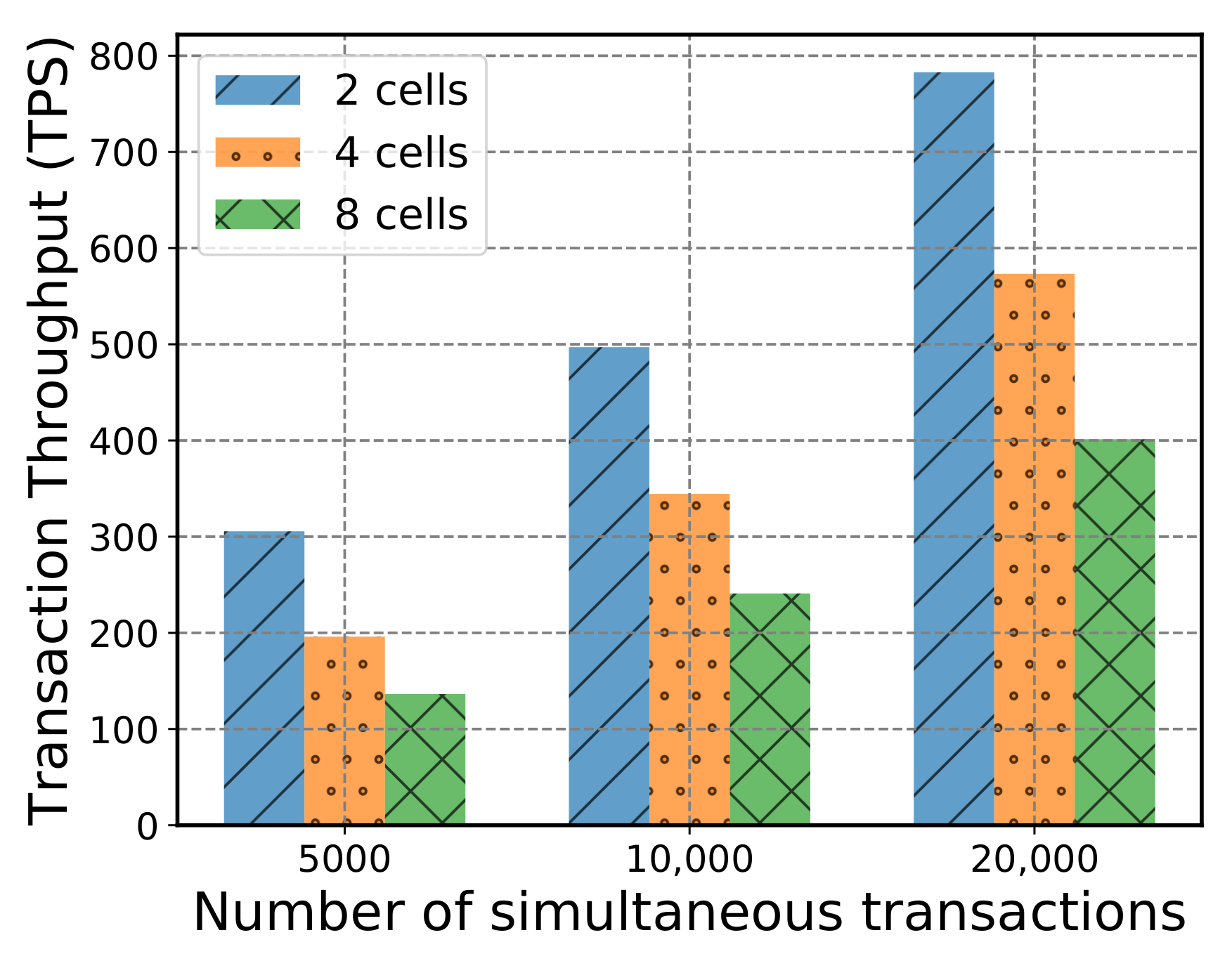}
    \caption{\textbf{Transaction throughput in \blockumulus.}}
    \label{fig:throughput}
\vspace{-5pt}
\end{figure}

For this evaluation, we transfer a small amount of funds from one \emph{FastMoney} account into another, measuring the full delay between the submission of the transaction until receiving the confirmation. We do not generate any failing transactions, nor do we observe any failures during the stress test. We run 9 experiments, matching three deployment configurations (2, 4, and 8 cells) with three sizes of transaction load (5,000, 10,000, and 20,000 simultaneous transactions), with the result shown in Fig.~\ref{fig:throughput}. The result demonstrates that: while the increased number of cells reduces the transaction throughput, the growing number of transactions makes the throughput larger, which is expected because the latency is growing slower than the number of simultaneous transactions, as was shown earlier. We attribute this ``bulk discount'' effect to the benefits of parallel execution, caching, and a significant reserve of available bandwidth due to the low communication overhead of \blockumulus.

\subsection{Operational Cost}
\blockumulus delivers transaction performance similar to credit card providers, alongside with decentralization properties seen in cryptocurrencies such as Bitcoin. This reconciliation of performance and decentralization comes at a price of delayed final settlement of transactions. Specifically, any confirmed transaction hinges upon trust towards the cells until the corresponding snapshot is submitted to the Ethereum smart contract. Therefore, the frequency of snapshot reports defines the speed of final irreversible settlement of recent transaction sets in \blockumulus. Table~\ref{tab:fees} shows how much each of the participating clouds will pay in Ethereum fees in 24 hours for data validation based on the frequency of the reports. Depending on the projected user participation and other goals of the \blockumulus deployment, the consortium can balance cost and frequency of reports. For comparison, the average price per Ethereum transaction on January 13, 2021 is \$5.72~\cite{ethfee}, with approximately 1,000 daily transactions~\cite{ntx}. With the same number of daily transactions, the \blockumulus fee overhead per transaction would be $218.08 / 1000 = \$0.218$ with 10-minute report frequency, which is about 26 times less than that in Ethereum. Moreover, the more subscribers a \blockumulus cell has, the lesser the amount of money is required per user to cover the reporting fee. For example, if a \blockumulus cell has 10,000 active subscribers, the monthly reporting fee overhead per user would be only \$0.65. We do not add the cost of auditing to the overall cost because cross-auditing is already a part of the normal cell operation, and the third-party auditing does not incur any expense for \blockumulus cell operators.

\begin{table}[]
    \centering
    \caption{\textbf{Cost of \blockumulus smart contract fees for participating cloud services based on the report period.}}
    \label{tab:fees}
    \begin{tabular}{|c|c|c|}
    \hline
    \textbf{Report}  & \multicolumn{2}{c|}{\textbf{Cost per 24 hours per cloud provider}} \\
    \cline{2-3}
    \textbf{Period} & \textbf{~~~~~~~~~~Gas~~~~~~~~~~} & \textbf{Approx. USD$^{\mathrm{a}}$} \\
    \hline\hline
    10 min & 7,083,792 & 218.08 \\
    \hline
    30 min & 2,361,264 & 72.69\\
    \hline
    1 hour & 1,180,632 & 36.35 \\
    \hline
    8 hours & 147,579 & 4.54 \\
    \hline
    24 hours & 49,193 & 1.51 \\
    \hline
    
    \multicolumn{3}{l}{$^{\mathrm{a}}$With the market price of Ether \$733 and gas price 22 GWei.}
    
    \end{tabular}
    
    \vspace{-20pt}
\end{table}

%\vspace{-5pt}
\section{Conclusion}\label{section:conclusion}
We propose \blockumulus, the first scalable framework for deploying decentralized smart contracts on the cloud, to address the blockchain scalability limitations on three dimensions: transaction throughput, data storage, and computation. The core idea of \blockumulus is to exploit a novel overlay consensus which delivers decentralization to smart contracts in a centralized cloud instead of random P2P network nodes. Concretely, a consortium of centralized cloud computing nodes can host a permissionless smart contract environment where clients can control the execution of their customized contracts and manage the data stored by these contracts. Our evaluation on Microsoft Azure shows that \blockumulus can execute tens of thousands of transactions within a minute, which is on par with the average throughput of worldwide credit card transactions. By integrating the decentralization of smart contracts and the scalability feature of the cloud, \blockumulus takes the first step towards high-performance data-rich smart contracts with high transaction throughput.
%We make \blockumulus open-source and invite researchers and engineers for participation and collaboration.

\section*{Acknowledgement}
We would like to thank our shepherd David Mohaisen  and anonymous reviewers for valuable feedback on our work. This work was supported in part by National Science Foundation grants CNS1950171, CNS1949753, and CNS2000681, as well as through Michigan State University’s Institute for Cyber-Enabled Research Cloud Computing Fellowship, with computational resources and services provided by Information Technology Services and the Office of Research and Innovation at Michigan State University.

%\vspace{-5pt}
\bibliographystyle{IEEEtran}

\begin{thebibliography}{10}
\providecommand{\url}[1]{#1}
\csname url@samestyle\endcsname
\providecommand{\newblock}{\relax}
\providecommand{\bibinfo}[2]{#2}
\providecommand{\BIBentrySTDinterwordspacing}{\spaceskip=0pt\relax}
\providecommand{\BIBentryALTinterwordstretchfactor}{4}
\providecommand{\BIBentryALTinterwordspacing}{\spaceskip=\fontdimen2\font plus
\BIBentryALTinterwordstretchfactor\fontdimen3\font minus
  \fontdimen4\font\relax}
\providecommand{\BIBforeignlanguage}[2]{{%
\expandafter\ifx\csname l@#1\endcsname\relax
\typeout{** WARNING: IEEEtran.bst: No hyphenation pattern has been}%
\typeout{** loaded for the language `#1'. Using the pattern for}%
\typeout{** the default language instead.}%
\else
\language=\csname l@#1\endcsname
\fi
#2}}
\providecommand{\BIBdecl}{\relax}
\BIBdecl

\bibitem{nakamoto2019bitcoin}
S.~Nakamoto, ``Bitcoin: A peer-to-peer electronic cash system,'' Manubot, Tech.
  Rep., 2019.

\bibitem{dembo2020everything}
A.~Dembo, S.~Kannan, E.~N. Tas, D.~Tse, P.~Viswanath, X.~Wang, and O.~Zeitouni,
  ``Everything is a race and nakamoto always wins,'' \emph{arXiv preprint
  arXiv:2005.10484}, 2020.

\bibitem{hardwick2018voting}
F.~S. Hardwick, A.~Gioulis, R.~N. Akram, and K.~Markantonakis, ``E-voting with
  blockchain: An e-voting protocol with decentralisation and voter privacy,''
  in \emph{2018 IEEE International Conference on Internet of Things
  (iThings)}.\hskip 1em plus 0.5em minus 0.4em\relax IEEE, 2018.

\bibitem{wang2018large}
B.~Wang, J.~Sun, Y.~He, D.~Pang, and N.~Lu, ``Large-scale election based on
  blockchain,'' \emph{Procedia Computer Science}, vol. 129, pp. 234--237, 2018.

\bibitem{ritzdorf2018tls}
H.~Ritzdorf, K.~W{\"u}st, A.~Gervais, G.~Felley, and S.~Capkun, ``Tls-n:
  Non-repudiation over tls enablign ubiquitous content signing.'' in
  \emph{NDSS}, 2018.

\bibitem{ivanov2020smart}
N.~Ivanov, J.~Lou, and Q.~Yan, ``Smart wifi: Universal and secure smart
  contract-enabled wifi hotspot,'' in \emph{International Conference on
  Security and Privacy in Communication Systems}.\hskip 1em plus 0.5em minus
  0.4em\relax Springer, 2020, pp. 425--445.

\bibitem{ramachandran2018smartprovenance}
A.~Ramachandran and M.~Kantarcioglu, ``Smartprovenance: a distributed,
  blockchain based dataprovenance system,'' in \emph{Proceedings of the Eighth
  ACM Conference on Data and Application Security and Privacy}, 2018, pp.
  35--42.

\bibitem{williams2016estonia}
O.~Williams-Grut, ``Estonia is using the technology behind bitcoin to secure 1
  million health records,'' \emph{Bus Insid}, 2016.

\bibitem{androulaki2018hyperledger}
E.~Androulaki, A.~Barger, V.~Bortnikov, C.~Cachin, K.~Christidis, A.~De~Caro,
  D.~Enyeart, C.~Ferris, G.~Laventman, Y.~Manevich \emph{et~al.}, ``Hyperledger
  fabric: a distributed operating system for permissioned blockchains,'' in
  \emph{Proceedings of the thirteenth EuroSys conference}, 2018, pp. 1--15.

\bibitem{cheng2019ekiden}
R.~Cheng, F.~Zhang, J.~Kos, W.~He, N.~Hynes, N.~Johnson, A.~Juels, A.~Miller,
  and D.~Song, ``Ekiden: A platform for confidentiality-preserving,
  trustworthy, and performant smart contracts,'' in \emph{2019 IEEE European
  Symposium on Security and Privacy (EuroS\&P)}.\hskip 1em plus 0.5em minus
  0.4em\relax IEEE, 2019.

\bibitem{poon2017plasma}
J.~Poon and V.~Buterin, ``Plasma: Scalable autonomous smart contracts,''
  \emph{White paper}, pp. 1--47, 2017.

\bibitem{wood2016polkadot}
G.~Wood, ``Polkadot: Vision for a heterogeneous multi-chain framework,''
  \emph{White Paper}, 2016.

\bibitem{zamani2018rapidchain}
M.~Zamani, M.~Movahedi, and M.~Raykova, ``Rapidchain: Scaling blockchain via
  full sharding,'' in \emph{Proceedings of the 2018 ACM SIGSAC Conference on
  Computer and Communications Security}, 2018, pp. 931--948.

\bibitem{gilad2017algorand}
Y.~Gilad, R.~Hemo, S.~Micali, G.~Vlachos, and N.~Zeldovich, ``Algorand: Scaling
  byzantine agreements for cryptocurrencies,'' in \emph{Proceedings of the 26th
  Symposium on Operating Systems Principles}, 2017, pp. 51--68.

\bibitem{poon2016bitcoin}
J.~Poon and T.~Dryja, ``The bitcoin lightning network: Scalable off-chain
  instant payments,'' 2016.

\bibitem{dziembowski2017perun}
S.~Dziembowski, L.~Eckey, S.~Faust, and D.~Malinowski, ``Perun: Virtual payment
  channels over cryptographic currencies.'' \emph{IACR Cryptol. ePrint Arch.},
  vol. 2017, p. 635, 2017.

\bibitem{biswas2019pobt}
S.~Biswas, K.~Sharif, F.~Li, S.~Maharjan, S.~P. Mohanty, and Y.~Wang, ``Pobt: A
  lightweight consensus algorithm for scalable iot business blockchain,''
  \emph{IEEE Internet of Things Journal}, vol.~7, no.~3, pp. 2343--2355, 2019.

\bibitem{sompolinsky2016spectre}
Y.~Sompolinsky, Y.~Lewenberg, and A.~Zohar, ``Spectre: A fast and scalable
  cryptocurrency protocol.'' \emph{IACR Cryptol. ePrint Arch.}, vol. 2016, p.
  1159, 2016.

\bibitem{chawla2019velocity}
N.~Chawla, H.~W. Behrens, D.~Tapp, D.~Boscovic, and K.~S. Candan, ``Velocity:
  Scalability improvements in block propagation through rateless erasure
  coding,'' in \emph{2019 IEEE International Conference on Blockchain and
  Cryptocurrency (ICBC)}.\hskip 1em plus 0.5em minus 0.4em\relax IEEE, 2019,
  pp. 447--454.

\bibitem{minsky1967computation}
M.~L. Minsky, \emph{Computation}.\hskip 1em plus 0.5em minus 0.4em\relax
  Prentice-Hall Englewood Cliffs, 1967.

\bibitem{zhou2020solutions}
Q.~Zhou, H.~Huang, Z.~Zheng, and J.~Bian, ``Solutions to scalability of
  blockchain: A survey,'' \emph{IEEE Access}, vol.~8, pp. 16\,440--16\,455,
  2020.

\bibitem{rfc2870}
\BIBentryALTinterwordspacing
D.~Karrenberg, M.~A. Kosters, R.~Plzak, and R.~Bush, ``{Root Name Server
  Operational Requirements},'' RFC 2870, Jun. 2000. [Online]. Available:
  \url{https://rfc-editor.org/rfc/rfc2870.txt}
\BIBentrySTDinterwordspacing

\bibitem{kwon2019impossibility}
Y.~Kwon, J.~Liu, M.~Kim, D.~Song, and Y.~Kim, ``Impossibility of full
  decentralization in permissionless blockchains,'' in \emph{Proceedings of the
  1st ACM Conference on Advances in Financial Technologies}, 2019.

\bibitem{decker2013information}
C.~Decker and R.~Wattenhofer, ``Information propagation in the bitcoin
  network,'' in \emph{IEEE P2P 2013 Proceedings}.\hskip 1em plus 0.5em minus
  0.4em\relax IEEE, 2013, pp. 1--10.

\bibitem{kalodner2018arbitrum}
H.~Kalodner, S.~Goldfeder, X.~Chen, S.~M. Weinberg, and E.~W. Felten,
  ``Arbitrum: Scalable, private smart contracts,'' in \emph{27th $\{$USENIX$\}$
  Security Symposium ($\{$USENIX$\}$ Security 18)}, 2018, pp. 1353--1370.

\bibitem{dai2019jidar}
X.~Dai, J.~Xiao, W.~Yang, C.~Wang, and H.~Jin, ``Jidar: A jigsaw-like data
  reduction approach without trust assumptions for bitcoin system,'' in
  \emph{2019 IEEE 39th International Conference on Distributed Computing
  Systems (ICDCS)}.\hskip 1em plus 0.5em minus 0.4em\relax IEEE, 2019, pp.
  1317--1326.

\bibitem{wang2019monoxide}
J.~Wang and H.~Wang, ``Monoxide: Scale out blockchains with asynchronous
  consensus zones,'' in \emph{16th $\{$USENIX$\}$ Symposium on Networked
  Systems Design and Implementation ($\{$NSDI$\}$ 19)}, 2019, pp. 95--112.

\bibitem{kokoris2018omniledger}
E.~Kokoris-Kogias, P.~Jovanovic, L.~Gasser, N.~Gailly, E.~Syta, and B.~Ford,
  ``Omniledger: A secure, scale-out, decentralized ledger via sharding,'' in
  \emph{2018 IEEE Symposium on Security and Privacy (SP)}.\hskip 1em plus 0.5em
  minus 0.4em\relax IEEE, 2018, pp. 583--598.

\bibitem{bowe2020zexe}
S.~Bowe, A.~Chiesa, M.~Green, I.~Miers, P.~Mishra, and H.~Wu, ``Zexe: Enabling
  decentralized private computation,'' in \emph{2020 IEEE Symposium on Security
  and Privacy (SP)}.\hskip 1em plus 0.5em minus 0.4em\relax IEEE, 2020, pp.
  947--964.

\bibitem{kiayias2017ouroboros}
A.~Kiayias, A.~Russell, B.~David, and R.~Oliynykov, ``Ouroboros: A provably
  secure proof-of-stake blockchain protocol,'' in \emph{Annual International
  Cryptology Conference}.\hskip 1em plus 0.5em minus 0.4em\relax Springer,
  2017, pp. 357--388.

\bibitem{yu2020ohie}
H.~Yu, I.~Nikoli{\'c}, R.~Hou, and P.~Saxena, ``Ohie: Blockchain scaling made
  simple,'' in \emph{2020 IEEE Symposium on Security and Privacy (SP)}.\hskip
  1em plus 0.5em minus 0.4em\relax IEEE, 2020, pp. 90--105.

\bibitem{ivanov2021targeting}
N.~Ivanov, J.~Lou, T.~Chen, J.~Li, and Q.~Yan, ``Targeting the weakest link:
  Social engineering attacks in ethereum smart contracts,'' in
  \emph{Proceedings of the 16th ACM Asia Conference on Computer and
  Communications Security}, 2021.

\bibitem{creditcard}
``The average number of credit card transactions per day \& year,''
  {https://www.cardrates.com/advice/number-of-credit-card-transactions-per-day-year/},
  accessed: 2021-01-12.

\bibitem{ethfee}
``Ethereum average transaction fee,''
  {https://ycharts.com/indicators/ethereum\_average\_transaction\_fee},
  accessed: 2021-01-13.

\bibitem{ntx}
``Ethereum daily transactions chart,'' {https://etherscan.io/chart/tx},
  accessed: 2021-01-13.

\end{thebibliography}
% Generated by IEEEtran.bst, version: 1.14 (2015/08/26)

\end{document}